\documentclass[aps,prl,twocolumn,superscriptaddress]{revtex4-1}
\usepackage{amssymb,amsmath}
\usepackage{graphicx}
\usepackage{dcolumn}
\usepackage{bm}
\usepackage{latexsym}
\usepackage{color}
\usepackage{times}
\usepackage{setspace}
\usepackage{hyperref}

\newcommand{\rfig}[1]{Fig.~\ref{#1}}

\newcommand{\req}[1]{Eq.~(\ref{#1})}

\hypersetup{colorlinks=true, linkcolor=blue, citecolor=blue, urlcolor=black}

\newcommand*{\red}{\textcolor{black}}   %%% for the 2th-revision
\newcommand*{\rred}{\textcolor{black}}    %%% for the 3th-revision

\begin{document}

\title{Simulation of Higher-Order Topological Phases and Related Topological Phase \\ Transitions in a Superconducting Qubit}
\author{Jingjing Niu}
\altaffiliation{These authors contributed equally to this work.}
\affiliation{Shenzhen Institute for Quantum Science and Engineering and Department of Physics, Southern University of Science and Technology, Shenzhen 518055, China}
\affiliation{Guangdong Provincial Key Laboratory of Quantum Science and Engineering, Southern University of Science and Technology, Shenzhen, 518055, China}
\affiliation{Shenzhen Key Laboratory of Quantum Science and Engineering, Southern University of Science and Technology, Shenzhen 518055, China}
\author{Tongxing Yan}
\altaffiliation{These authors contributed equally to this work.}
\affiliation{Shenzhen Institute for Quantum Science and Engineering and Department of Physics, Southern University of Science and Technology, Shenzhen 518055, China}
\affiliation{Guangdong Provincial Key Laboratory of Quantum Science and Engineering, Southern University of Science and Technology, Shenzhen, 518055, China}
\affiliation{Shenzhen Key Laboratory of Quantum Science and Engineering, Southern University of Science and Technology, Shenzhen 518055, China}
\author{Yuxuan Zhou}
\altaffiliation{These authors contributed equally to this work.}
\affiliation{Shenzhen Institute for Quantum Science and Engineering and Department of Physics, Southern University of Science and Technology, Shenzhen 518055, China}
\affiliation{Guangdong Provincial Key Laboratory of Quantum Science and Engineering, Southern University of Science and Technology, Shenzhen, 518055, China}
\affiliation{Shenzhen Key Laboratory of Quantum Science and Engineering, Southern University of Science and Technology, Shenzhen 518055, China}
\author{Ziyu Tao}
\affiliation{Shenzhen Institute for Quantum Science and Engineering and Department of Physics, Southern University of Science and Technology, Shenzhen 518055, China}
\affiliation{Guangdong Provincial Key Laboratory of Quantum Science and Engineering, Southern University of Science and Technology, Shenzhen, 518055, China}
\affiliation{Shenzhen Key Laboratory of Quantum Science and Engineering, Southern University of Science and Technology, Shenzhen 518055, China}
\author{Xiaole Li}
\affiliation{Shenzhen Institute for Quantum Science and Engineering and Department of Physics, Southern University of Science and Technology, Shenzhen 518055, China}
\affiliation{Guangdong Provincial Key Laboratory of Quantum Science and Engineering, Southern University of Science and Technology, Shenzhen, 518055, China}
\affiliation{Shenzhen Key Laboratory of Quantum Science and Engineering, Southern University of Science and Technology, Shenzhen 518055, China}
\author{Weiyang Liu}
\affiliation{Shenzhen Institute for Quantum Science and Engineering and Department of Physics, Southern University of Science and Technology, Shenzhen 518055, China}
\affiliation{Guangdong Provincial Key Laboratory of Quantum Science and Engineering, Southern University of Science and Technology, Shenzhen, 518055, China}
\affiliation{Shenzhen Key Laboratory of Quantum Science and Engineering, Southern University of Science and Technology, Shenzhen 518055, China}
\author{\\Libo Zhang}
\affiliation{Shenzhen Institute for Quantum Science and Engineering and Department of Physics, Southern University of Science and Technology, Shenzhen 518055, China}
\affiliation{Guangdong Provincial Key Laboratory of Quantum Science and Engineering, Southern University of Science and Technology, Shenzhen, 518055, China}
\affiliation{Shenzhen Key Laboratory of Quantum Science and Engineering, Southern University of Science and Technology, Shenzhen 518055, China}
\author{Hao Jia}
\affiliation{Shenzhen Institute for Quantum Science and Engineering and Department of Physics, Southern University of Science and Technology, Shenzhen 518055, China}
\affiliation{Guangdong Provincial Key Laboratory of Quantum Science and Engineering, Southern University of Science and Technology, Shenzhen, 518055, China}
\affiliation{Shenzhen Key Laboratory of Quantum Science and Engineering, Southern University of Science and Technology, Shenzhen 518055, China}
\author{Song Liu}
\thanks{lius3@sustech.edu.cn}
\affiliation{Shenzhen Institute for Quantum Science and Engineering and Department of Physics, Southern University of Science and Technology, Shenzhen 518055, China}
\affiliation{Guangdong Provincial Key Laboratory of Quantum Science and Engineering, Southern University of Science and Technology, Shenzhen, 518055, China}
\affiliation{Shenzhen Key Laboratory of Quantum Science and Engineering, Southern University of Science and Technology, Shenzhen 518055, China}
\author{Zhongbo Yan}
\thanks{yanzhb5@mail.sysu.edu.cn}
\affiliation{School of Physics, Sun Yat-Sen University, Guangzhou 510275, China}
\author{Yuanzhen Chen}
\thanks{chenyz@sustech.edu.cn}
\affiliation{Shenzhen Institute for Quantum Science and Engineering and Department of Physics, Southern University of Science and Technology, Shenzhen 518055, China}
\affiliation{Guangdong Provincial Key Laboratory of Quantum Science and Engineering, Southern University of Science and Technology, Shenzhen, 518055, China}
\affiliation{Shenzhen Key Laboratory of Quantum Science and Engineering, Southern University of Science and Technology, Shenzhen 518055, China}

\author{Dapeng Yu}
\affiliation{Shenzhen Institute for Quantum Science and Engineering and Department of Physics, Southern University of Science and Technology, Shenzhen 518055, China}
\affiliation{Guangdong Provincial Key Laboratory of Quantum Science and Engineering, Southern University of Science and Technology, Shenzhen, 518055, China}
\affiliation{Shenzhen Key Laboratory of Quantum Science and Engineering, Southern University of Science and Technology, Shenzhen 518055, China}

%\date{\today}

\begin{abstract}
Higher-order topological phases  give rise to new bulk and boundary physics, as well as new classes of topological phase transitions. \red{While the realization of higher-order topological phases has been confirmed in many platforms by detecting the existence of gapless boundary modes, a direct determination of the higher-order topology and related topological phase transitions through the bulk in experiments has still been lacking.}
\red{To bridge the gap}, in this work we carry out the simulation of a two-dimensional second-order \red{topological phase} in a superconducting qubit. Owing to the great flexibility and controllability of the quantum simulator, we  observe the realization of higher-order topology directly through the measurement of the pseudo-spin texture in momentum space of the bulk for the first time, in sharp contrast to previous experiments based on the detection of  gapless boundary modes in real space. Also through the measurement of the evolution of pseudo-spin texture with parameters, we further observe novel topological phase transitions from the second-order \red{topological phase} to the trivial \red{phase}, as well as to the first-order \red{topological phase} with nonzero Chern number. Our work sheds new light on the study of higher-order topological phases and  topological phase transitions.\vspace{10pt}
\end{abstract}
%\pacs{}

\maketitle

%\section{Introduction}
\begin{spacing}{2.0}
\noindent\textbf{1\quad Introduction}
\end{spacing}

The bulk-boundary correspondence is a fundamental principle of topological phases of matter. Very recently, higher-order \red{topological
insulators} (TIs) and \red{topological superconductors} (TSCs) have attracted broad interest owing to their unconventional bulk-boundary correspondence ~\cite{Benalcazar2017,Schindler2018,Song2017,Langbehn2017,Benalcazar2017a,Ezawa2018,Khalaf2018a,Geier2018,Franca2018,Trifunovic2019}. In comparison to their conventional counterparts, also known as first-order TIs and TSCs~\cite{Hasan2010,Qi2011}, the unconventionality is manifested through the codimension of their gapless boundary modes. Concretely,  the boundary modes of $n$th\rred{-}order TIs or TSCs have codimension $n$, with $n=1$ and $n\geq2$ corresponding to the first-order and higher-order ones, respectively.

Two- and three-dimensional higher-order TIs have already been experimentally realized in many platforms, including photonic crystals~\cite{noh2018topological,Chen2019photonic,Xie2108photonic,Hassan2019corner}, microwave resonators~\cite{peterson2018quantized}, electric circuits ~\cite{imhof2018corner,Bao2019octupole}, phononic metamaterials~\cite{serra2018observation,xue2019acoustic,zhang2019second,Xue2019octupole,Ni2019octupole}, and a few electronic materials~\cite{Schindler2018HOTI,Kempkes2019}. In \red{comparison}, higher-order TSCs have so far been little explored in experiments~\cite{Gray2019helical}, owing to the underlying difficulty in realizing this class of novel phases in real materials~\cite{Zhu2018,Yan2018,Wang2018hosc,Wang2018hosc2,Liu2018hosc,Hsu2018,Wu2019hosc,Zhang2019hinge,
Volpez2019SOTSC,Zhu2019SOTSC,Franca2019SOTSC,Yang2019hinge,Ghorashi2019HOSC,Pan2018SOTSC,Zhang2019hoscb,
Yan2019second,Wu2019hoscb,Hsu2019HOSC,Wu2019swave}. \red{Similar to the first-order topology, the higher-order
topology is defined by the bulk momentum-space Hamiltonian. However, thus far the determination of
higher-order topology in experiments has been indirect and relying on the detection of gapless
modes at the theoretically predicted positions on the real-space boundary.
Moreover,  experimental works on higher-order topological phases have also
mainly focused on the gapless boundary modes, novel physics directly related to the bulk,
like topological phase transitions, has still been little explored in experiments~\cite{Serra2019quadrupole}.
Particularly, while the higher-order topological phases bring new possibility
to topological phase transitions,
we notice  that the novel class, which take place between higher-order
and first-order topological phases within the same symmetry class, have yet to be  investigated experimentally.}

\red{Although
the indirect boundary approach is simple in experiments and the bulk-boundary correspondence guarantees its reliability,
a direct bulk approach, if possible, is highly desirable as it can directly determine the underlying
topological invariants and thus can detect topological phase transitions much more precisely than
the boundary approach. In real materials, it is obvious that the bulk approach is rather challenging
because the complexity of band structures is high and the topological invariants depend on
all occupied bands\cite{Chiu2016review}. Nevertheless, as the essential physics of various
topological phases can also be realized in some simple Hamiltonians which only involve
a minimal set of bands, the great reduction of complexity in such situations will
make the bulk approach feasible.
For instance, when the concerned system is described by a two-band Hamiltonian of the form
 $H(\mathbf{k})=\mathbf{d}(\mathbf{k})\cdot \boldsymbol{\sigma}$ with $\boldsymbol{\sigma}=(\sigma_{x},\sigma_{y},\sigma_{z})$ the Pauli matrices, its topological property can be simply determined by measuring the spin (or  pseudo-spin in general) texture throughout the Brillouin zone (BZ)~\cite{Roushan2014,Schroer2014TPT,Flurin2017,Xu2018winding}. One celebrated example is the Qi-Wu-Zhang model~\cite{Qi2006model}, which describes a Chern insulator when the spin texture realizes a Skyrmion configuration in the BZ. }

In this work,  we carry out the simulation of \red{a two-dimensional two-band Hamiltonian which can realize both
second-order and first-order topological  phases} in a superconducting qubit. By mapping the momentum space of the simulated two-band Hamiltonian to the parameter space of the qubit Hamiltonian, we are able to determine the pseudo-spin texture in the whole BZ with a combinational use of quantum-quench dynamics and quantum state tomography (QST). Through the evolution of pseudo-spin texture with parameters, we not only observe  the topological phase transitions between second-order \red{topological phases} and trivial \red{phases}, but also observe the ones between second-order and first-order \red{topological phases} within the same symmetry class for the first time.

\begin{spacing}{2.0}
\noindent\textbf{2\quad Results}
\begin{spacing}{1.5}
\noindent\it{2.1 Theoretical model.}
\end{spacing}
\end{spacing}
\red{In terms of the Pauli matrices,  an arbitrary two-band Hamiltonian can be written as
$H=\sum_{\mathbf{k}}\psi_{\mathbf{k}}^{\dag}H(\mathbf{k})\psi_{\mathbf{k}}$,
where $\psi_{\mathbf{k}}$ denotes a basis of two degrees of freedom, and
\begin{eqnarray}
H(\mathbf{k})=d_{0}(\mathbf{k})\mathbf{I}_{2\times2}+\sum_{i=x,y,z}d_{i}(\mathbf{k})\sigma_{i}.\label{BdG}
\end{eqnarray}
As the first term with two-by-two identity  matrix $\mathbf{I}_{2\times2}$ plays no role
in the band topology, we will let it vanish throughout this work. }

In a recent work, one of us revealed that when the $\mathbf{d}=(d_{x},d_{y},d_{z})$ vector is constructed by a Hopf map, the resulting Hamiltonian provides  a minimal-model realization of second-order \red{topological  phases}~\cite{Yan2019HOTOPSC}. Concretely, the Hopf map is $d_{i}(\mathbf{k})=z^{\dag}(\mathbf{k})\sigma_{i}z(\mathbf{k})$, where the spinor $z(\mathbf{k})=(z_{1}(\mathbf{k}),z_{2}(\mathbf{k}))^{T}$,  $z_{1}(\mathbf{k})=f_{1}(\mathbf{k})+if_{2}(\mathbf{k})$, and
$z_{2}(\mathbf{k})=g_{1}(\mathbf{k})+ig_{2}(\mathbf{k})$, with $f_{1}(\mathbf{k})=(\cos k_{x}+\lambda_{1})$, $f_{2}(\mathbf{k})=(\cos k_{y}+\lambda_{2})$,
$g_{1}(\mathbf{k})=\sin k_{x}$, $g_{2}(\mathbf{k})=\sin k_{y}$. Accordingly, $d_{x}=2(f_{1}g_{1}+f_{2}g_{2})$,
$d_{y}=2(f_{1}g_{2}-f_{2}g_{1})$, and $d_{z}=f_{1}^{2}+f_{2}^{2}-g_{1}^{2}-g_{2}^{2}$.
Remarkably, this model has a simple phase diagram resembling the one of the Benalcazar-Bernevig-Hughes model~\cite{Benalcazar2017}. That is, when $|\lambda_{1,2}|<1$, it realizes a second-order \red{topological phase}, otherwise  it describes a \red{trivial  phase}. However, a fundamental difference lies between this model and the Benalcazar-Bernevig-Hughes model. For the latter, it belongs to either the class BDI or the class AI (depending on whether an on-site potential is present or not) of the Altland-Zirnbauer classification~\cite{Schnyder2008,Kitaev2009,Ryu2010}. In two dimensions, it is known that both class BDI and class AI do not allow any first-order topological phases.  In contrast, the model given in Ref.~\cite{Yan2019HOTOPSC} belongs to either the class D \red{or the class A (depending on whether the term $d_{0}(\mathbf{k})\mathbf{I}_{2\times2}$,
which will break the particle-hole symmetry, is present or not)}.
In two dimensions, it is known that both class D and class A allow first-order topological phases
which are characterized by the first-class Chern number~\cite{qi2010chiral}.
As second-order and first-order topological phases are both allowed in the class D \red{and class A, this raises the possibility to observe
topological phase transitions between topological phases of different orders.}

Based on the above recognition, we lift the strong constraint imposed by the Hopf map by introducing a new free parameter
to the model given in Ref.~\cite{Yan2019HOTOPSC}. For concreteness, we write down $d_{x,y,z}$ explicitly, which read
\begin{small}
\begin{eqnarray}
d_{x}(\mathbf{k})&=&2\lambda_1 \sin k_x + 2\lambda_2 \sin k_y + \sin 2k_x +\sin 2k_y,\nonumber\\
d_{y}(\mathbf{k})&=&2\lambda_1 \sin k_y - 2\lambda_2 \sin k_x + 2\sin (k_y-k_x),\nonumber\\
d_{z}(\mathbf{k})&=&2\lambda_1 \cos k_x + 2\lambda_2 \cos k_y + \cos 2k_x +\cos 2k_y\!-\!\mu,\label{expression}
\end{eqnarray}
\end{small}
where $\mu$ is the newly-added free parameter. If we fix $\mu=-(\lambda_{1}^{2}+\lambda_{2}^{2})$, the above Hamiltonian reduces to the one in Ref.~\cite{Yan2019HOTOPSC}. As we will show shortly, this generalized Hamiltonian has a richer phase diagram, most importantly, it allows topological phase transitions between second-order and first-order topological phases.

\begin{figure}[htbp]
	\begin{center}
		\includegraphics[width=0.45\textwidth]{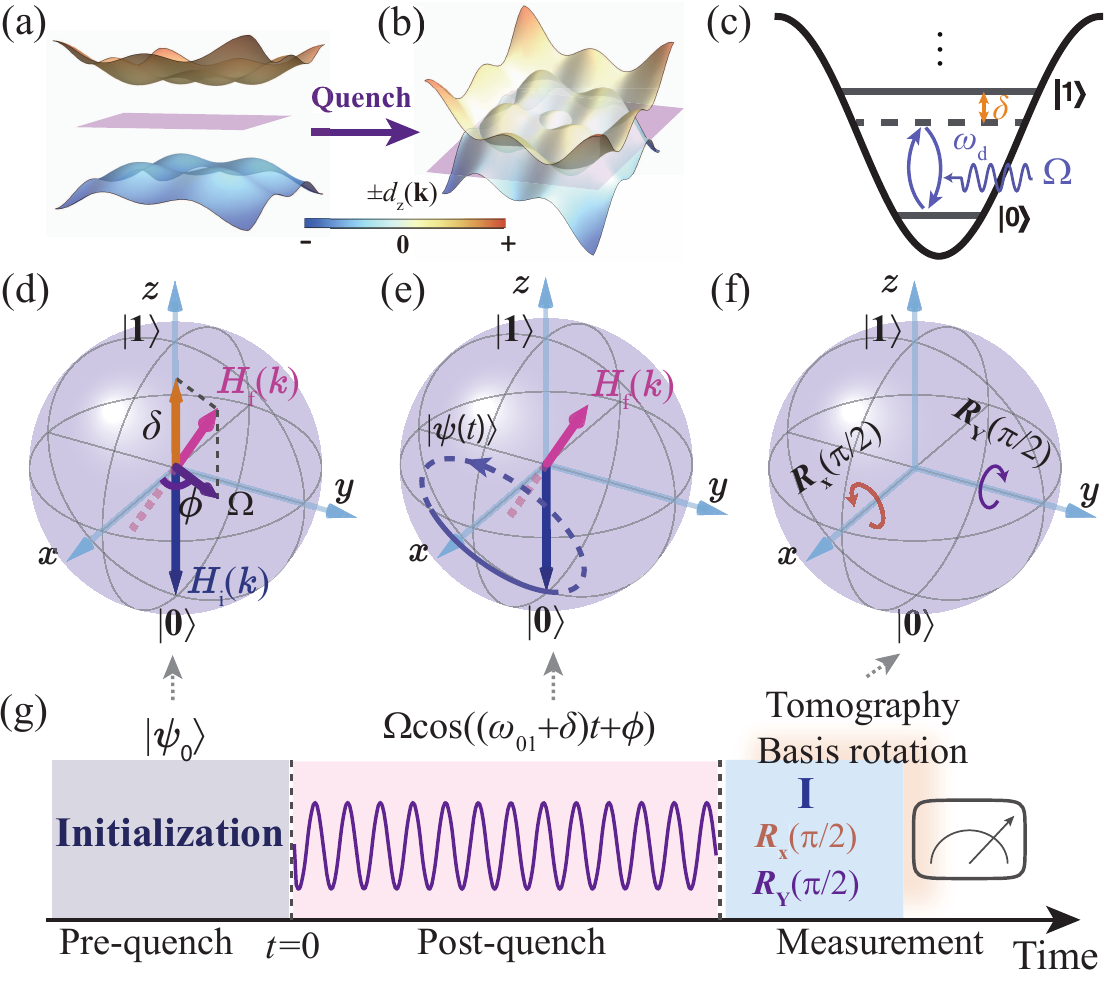}
		\caption{Schematics of principle and experimental implementation. $\pm d_{z}(\mathbf{k})$, whose crossings correspond to
the band inversion surfaces, for (a) pre-quench ($\lambda \!\rightarrow \! \infty$) and (b) post-quench ($\lambda\!=\!-0.5$) cases. (c) The energy diagram of an Xmon qubit. A microwave pulse is applied for simulation at each momentum via controlling its detuning frequency $\delta$, amplitude $\Omega$, and phase $\phi$. (d) corresponds to the moment right upon the application of the quench pulse that changes the initial Hamiltonian \rred{$H\rm_i$} to the final Hamiltonian \rred{$H\rm_f$}. (e) illustrates the temporal dynamics afterwards. (f) shows rotation pulses for the QST process. (g) Time line of experimental operations.}
		\label{Fig1}
	\end{center}
\end{figure}

\red{While the existence of particle-hole symmetry allows the above Hamiltonian to describe
either an insulator or a spinless \rred{superconductor},
in this work we will not emphasize the interpretation of the Hamiltonian
 because the experiment is performed in a single
superconducting qubit. Nevertheless, the bulk topology does not depend on the interpretation, so
we can still follow the analysis in Ref.~\cite{Yan2019HOTOPSC}. That is,
 as the two-band Hamiltonian has inversion symmetry ($\mathcal{I}H(\mathbf{k})\mathcal{I}^{-1}=H(-\mathbf{k})$ with $\mathcal{I}=\sigma_{z}$), its topological property is simply determined by the relative configuration between $d_{z}(\mathbf{k})=0$ (here dubbed band inversion surface (BIS) \rred{\cite{Zhang2018DTPT}}) and
$d_{x}(\mathbf{k}) = d_{y}(\mathbf{k}) = 0$ (dubbed Dirac points (DPs))}.
As the $d_{z}(\mathbf{k})$ term consists of both the nearest-neighbor and the next-nearest-neighbor hoppings, the number of \red{BISs ($N_{\rm BIS}$)}, which counts the number of disconnected contours satisfying
$d_{z}(\mathbf{k})=0$, can be $0$, $1$, and $2$. For the first-order topology,
the parity of Chern number $C$ is directly tied to \red{$N_{\rm BIS}$, namely $(-1)^{C}=(-1)^{N_{\rm BIS}}$}~\cite{sato2010odd}, indicating that a gapped phase with an odd number of \red{BIS} must have nontrivial first-order topology. To realize second-order topological phases,
it was revealed in Ref.~\cite{Yan2019HOTOPSC} that the number of \red{BIS} needs to
be even and removable \red{DPs} (not pinned at any specific momentum) are required
to be present between or within the disconnected \red{BISs} so that the resulting gapped phases cannot be continuously deformed to the trivial phase (no BIS) without the closure of bulk gap.

\vspace{10pt}
\begin{spacing}{1.5}
\noindent\it{2.2 Quench dynamics}
\end{spacing}
In this work, we simulate the two-band Hamiltonian (\req{BdG}) in an Xmon type superconducting qubit (see Supplemental Information (SI) for details of samples and experimental setup) and adopt QST to determine the underlying pseudo-spin texture. Concretely, we first prepare the qubit to stay in \rred{$|\psi_\mathrm{i}(\mathbf{k},t)\rangle$}, the eigenstate of $\sigma_{z}$, i.e., \rred{$\sigma_{z}|\psi_\mathrm{i}(\mathbf{k},t)\rangle=-|\psi_\mathrm{i}(\mathbf{k},t)\rangle$}. Such a choice corresponds to the ground state of $H(\mathbf{k})$ in the limit $\lambda_{1,2}\!\rightarrow\!\infty$ for which the Hamiltonian is trivial in topology (see \rfig{Fig1}(a)). Next, we suddenly quench the system at $t=0$ by a microwave pulse $\Omega(t) = \Omega\cos((\omega_{01}+\delta)t+\phi)$, with amplitude $\Omega\!=\!\sqrt{d_x^2+d_y^2}$, phase $\phi\!=\!\arctan{(d_y/d_x)}$, and frequency detuning $\delta\!=\!d_z$ (see \rfig{Fig1}(c)(d)), then the state will follow a unitary time evolution, i.e., \rred{$|\psi_\mathrm{f}(\mathbf{k},t)\rangle=e^{i H_\mathrm{f}(\mathbf{k})t}|\psi_\mathrm{i}(\mathbf{k},0)\rangle$}, where $\rred{H\!\rm_{f}(\mathbf{k})}=\mathbf{d}(\mathbf{k})\cdot\boldsymbol{\sigma}$ takes the form we desire to simulate.

\begin{figure}[htbp]
	\begin{center}
		\includegraphics[width=0.45\textwidth]{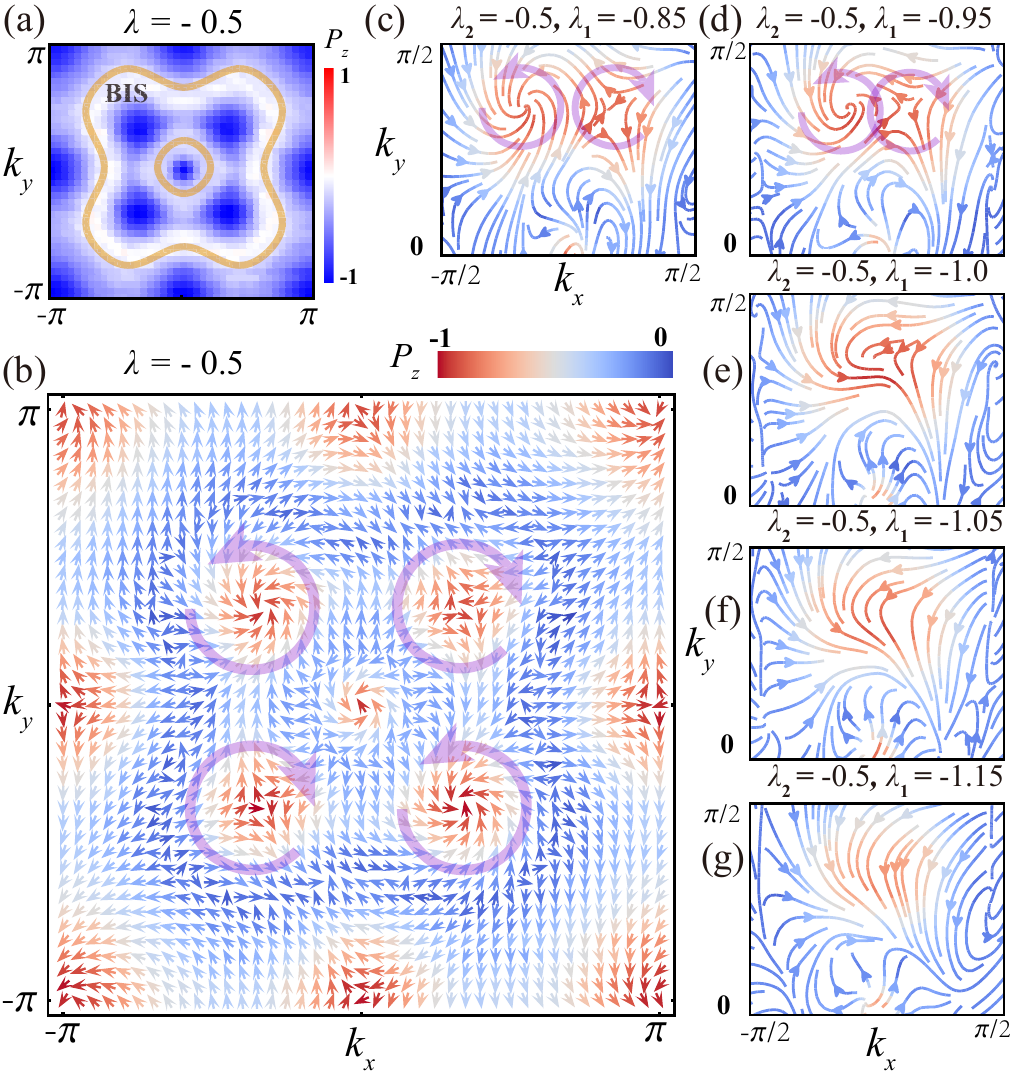}
		\caption{Experiments demonstrating a second-order \red{topological phase} and the movement and annihilation of (anti)vortices. (a) $P_z$ measured across the BZ for a particular realization of the second-order \red{topological phase} ($\lambda=-0.5$). The two closed curves indicate the \red{BISs} in the first BZ where $P_z=0$. (b) ($P_x$, $P_y$) measured across the BZ for the same realization, shown as two dimensional vectors. For completeness, $P_z$ is also indicated by the color of the vectors. The circular arrows mark the four removable vortices and antivortices. (c)-(g) The movement, merging, and annihilation of a pair of vortex and antivortex manifested by evolution of the pseudo-spin texture ($P_x$, $P_y$) for $\lambda_2=-0.5$, $\lambda_1=-1+\delta$, with $\delta\in[0.15,0.05,0,-0.05,-0.15]$.
                }
    \label{Fig2}
	\end{center}
\end{figure}

After the quench, the pseudo-spin polarization, defined as \rred{$\mathbf{P}(\mathbf{k},t)=\langle \psi_{f}(\mathbf{k},t)|\boldsymbol{\sigma}|\psi_{f}(\mathbf{k},t)\rangle$}, will precess on the Bloch sphere (see illustration in \rfig{Fig1}) around the direction of $\mathbf{d}(\mathbf{k})$. In experiments, the evolution of $\mathbf{P}(\mathbf{k}, t)$ can be measured by QST. Interestingly, it was shown in Ref.~\cite{Zhang2018DTPT} that the time-averaged pseudo-spin polarization is directly related to $\mathbf{d}(\mathbf{k})$ and then the band topology of the Hamiltonian can be extracted from this quantity\cite{Zhang2018DTPT,Sun2018a,Wang2019b,Yi2019a,Zhang2019b,Hu2019}. To obtain this quantity, here we flatten the Hamiltonian as \rred{$H_\mathrm{f}(\mathbf{k})/E_\mathrm{f}(\mathbf{k})$} with \rred{$E_\mathrm{f}(\mathbf{k})$} the eigenenergy of \rred{$H_\mathrm{f}(\mathbf{k})$} (such a procedure does not change the underlying topological properties), so that the period of evolution is identical for each \red{$\mathbf{k}$}~\cite{Hu2019,Guo2019}.  By doing so, we find that only two periods are sufficient to obtain a trustworthy value of the time-averaged pseudo-spin polarization (see SI for a detailed discussion). The result reads
\begin{eqnarray}\label{Px}
P_{i}(\mathbf{k})=-\frac{d_{i}(\mathbf{k})d_z(\mathbf{k})}{d_x^2(\mathbf{k})+d_y^2(\mathbf{k})+d_z^2(\mathbf{k})},
\end{eqnarray}
where $P_{i}$ with $i=x,y,z$ represent the three components (as shown in Figs. \ref{Fig1}(d-f)). It is immediately seen that the \red{BISs} determined by $d_{z}\!=\!0$ correspond to $P_z\!=\!0$, and the \red{DPs} determined by  $d_{x}\!=\!d_{y}\!=\!0$ can be extracted from $P_x\!=\!P_{y}\!=\!0$ after the contour $P_z\!=\!0$ is determined.

\begin{figure*}[htbp]
	\begin{center}
		\includegraphics[width=0.8\textwidth]{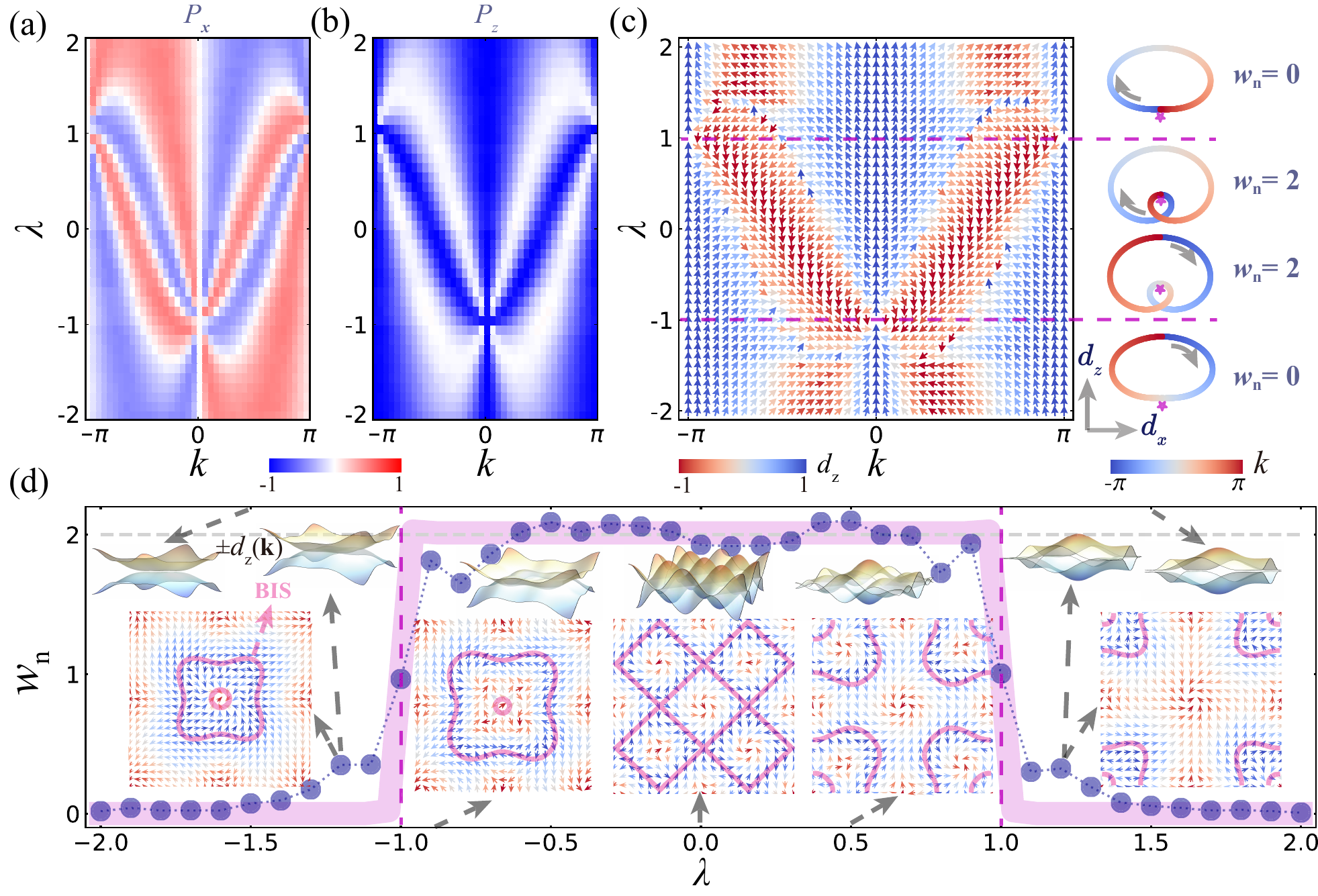}
		\caption{Phase transition from a second-order \red{topological phase} to a trivial \red{phase}. \red{Under a uniform discretization with $41$ discrete momentum points in $k\in[-\pi,\pi]$}, (a) and (b) respectively show the $P_x$ and $P_z$ measured along the $k_x=k_y$ line as $\lambda_1=\lambda_2=\lambda$ changes continually from $-2$ to $2$. (c) Left: $(d_{x},d_{z})$ versus $k$ for different $\lambda$ deduced from data in (a) and (b). On the right side the winding configuration of ($d_x$, $d_z$) around the origin (the stars) is shown, and the corresponding $\mathcal{W}_\text{n}$ is indicated. (d) $\lambda$-dependence of $\mathcal{W}_\text{n}$ obtained experimentally using \req{Wn} (blue circles) and the theoretical expectation (pink line). The insets show \red{$\pm d_{z}(\mathbf{k})$ (their crossings occur
at $d_{z}(\mathbf{k})=0$, corresponding to
the BISs)}, pseudo-spin texture ($P_x$, $P_z$), as well as the \red{BISs} ($P_z$ = 0), for different phases.}
		\label{Fig3}
	\end{center}
\end{figure*}

\begin{figure*}[htbp]
	\begin{center}
		\includegraphics[width=0.85\textwidth]{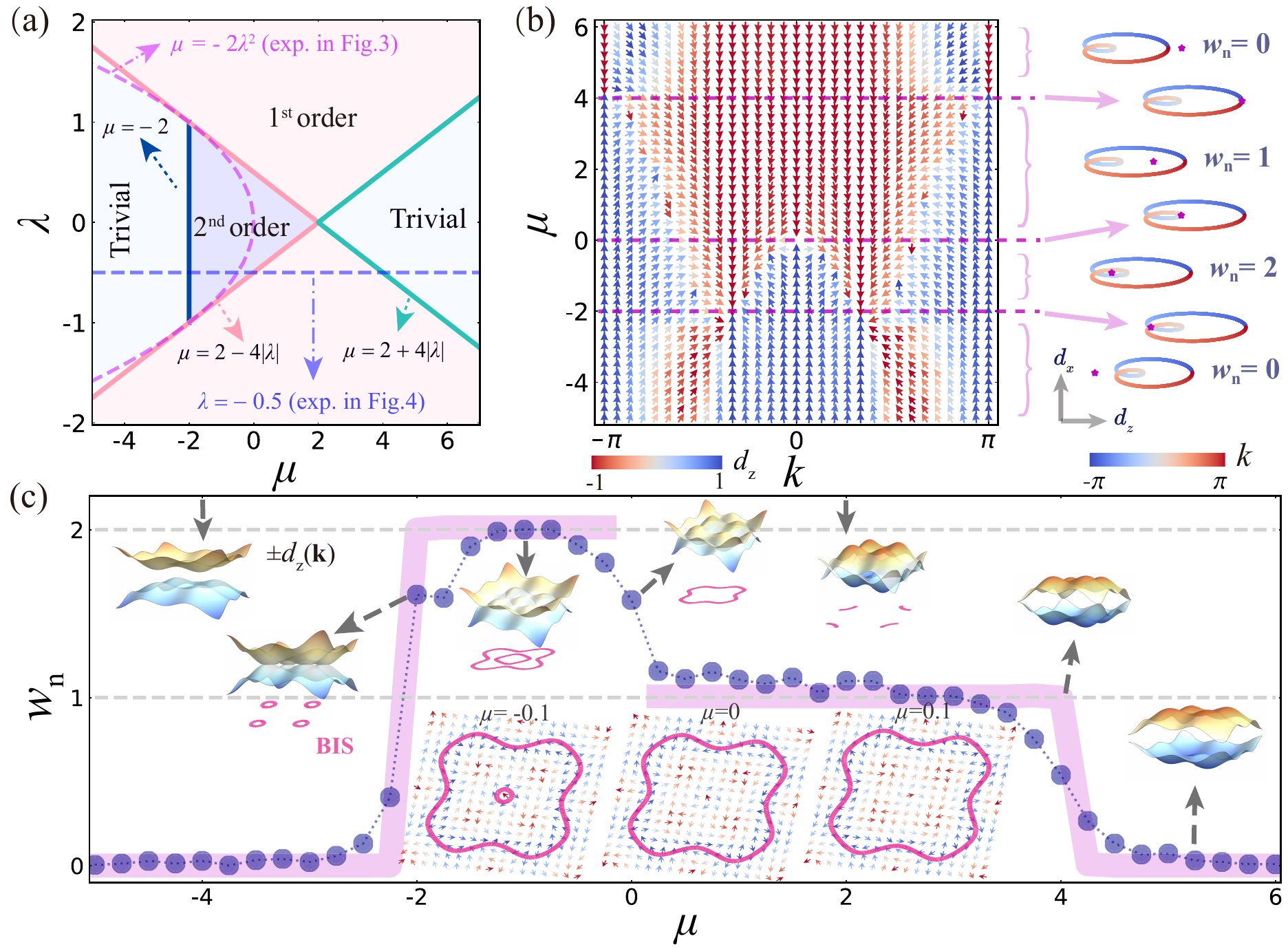}
		\caption{Phase transitions between first- and second-order \red{topological phases} and trivial \red{phases}. (a) Phase diagram of the simulated Hamiltonian with $\lambda_1=\lambda_2=\lambda$. The blue dashed line is for $\lambda=-0.5$, which is experimentally demonstrated in (c), while the pink curve represents the experiment in \rfig{Fig3}.  \red{Under a uniform discretization with $31$ discrete momentum points in $k\in[-\pi,\pi]$,} (b) shows the $(d_{x},d_{z})$ measured along the $k_{x}=k_{y}=k$ line for different $\mu$ with $\lambda$ fixed to $-0.5$. The right side shows the corresponding winding configurations ($d_z$, $d_x$). (c) $\mu$-dependence of $\mathcal{W}_\text{n}$ obtained experimentally using \req{Wn} (blue circles) and the theoretical expectation (pink line). The insets show \red{$\pm d_{z}(\mathbf{k})$}, pseudo-spin texture ($P_x$, $P_y$), as well as the \red{BISs} for different phases.}
		\label{Fig4}
	\end{center}
\end{figure*}

\vspace{10pt}
\begin{spacing}{1.5}
\noindent\it{2.3 Topological phase transitions between second-order topological phases and trivial phases}
\end{spacing}
We first show the experimental results for the case with $\mu$ fixed to $-(\lambda_1^2+\lambda_{2}^2)$, accordingly, the Hamiltonian hosts only two possible phases, a second-order \red{topological phase} and a trivial \red{phase}~\cite{Yan2019HOTOPSC}. Let us focus on some specific points in the parameter space at first. As shown in \rfig{Fig2}(a), when we set $\lambda_{1}\!=\!\lambda_{2}\! = \!-0.5$, there are two disconnected contours that satisfy $P_{z}=0$, indicating the presence of two disconnected \red{BISs} in the BZ. Figure \ref{Fig2}(b) shows the corresponding texture of $(P_{x},P_{y})$. It is readily seen that there are four vortices and four antivortices in the BZ whose cores correspond to $P_{x}=P_{y}=0$, with half of them located at time-reversal invariant momenta, and the other half located at some generic momenta between the two contours for $P_{z}=0$. As mentioned previously, these vortices and antivortices refer to \red{DPs}, and the four at generic momenta represent the removable ones which are crucial for the realization of the second-order topological phase~\cite{Yan2019HOTOPSC}. By fixing $\lambda_{2}=-0.5$ and decreasing $\lambda_{1}$, we find that the removable vortices and antivortices move toward each other in pair and then annihilate at $\lambda_{1}=-1$, as shown in Figs. \ref{Fig2}(c-g). After the annihilation, the two contours for $P_{z}=0$ can continuously move together and then annihilate without crossing any other vortices, indicating that the annihilation of removable vortices and antivortices corresponds to a topological phase transition from the second-order \red{topological phase} to the trivial \red{phase}.

To further confirm the topological phase transition, we tune the parameter to satisfy $\lambda_{1}=\lambda_{2}=\lambda$. Under this condition, the topological property of
the simulated Hamiltonian is fully characterized by the winding number defined on
the line $k_{x}=k_{y}$ or $k_{x}=-k_{y}$ (they are equivalent)~\cite{Yan2019HOTOPSC}.
Focusing on the line $k_{x}=k_{y}=k$ (see SI for the $k_x\!=\!-k_y$ case), the winding number is given by
\begin{eqnarray}
        \mathcal{W}_\text{n} = \frac{1}{2\pi}\int_{-\pi}^{\pi} \frac{d_z\partial_{k} d_x - d_x \partial_{k} d_z}{d_x^2+d_z^2}\, dk. \label{Wn}
\end{eqnarray}
\red{It is noteworthy that the above formula is an integral, however, a continuous measurement
of \rred{pseudo-}spin textures is apparently unrealistic in any real experiment.
While an intensive discretization of the BZ can reduce the errors caused by the finite
discretization as well as some random errors presented in the measurements of pseudo-spin polarization (see SI
for more discussion),
this needs to be balanced with a reasonable measurement time, otherwise it becomes difficult to explore as much
of the parameter space as possible. Under a uniform discretization with $41$ discrete momentum points in $k\in[-\pi,\pi]$,}
Figs. \ref{Fig3}(a) and (b) show the measured ($P_{x}$, $P_{z}$) through which the dependence of ($d_{x}$, $d_{z}$), and so $\mathcal{W}_\text{n}$, on $\lambda$, is extracted, as shown in \rfig{Fig3}(c). Geometrically, $\mathcal{W}_\text{n}$ corresponds to the number of cycles that the vector ($d_{x}$, $d_{z}$) winds around the origin ($d_{x}=0, d_{z}=0$) when $k$ varies from $-\pi$ to $\pi$~\cite{Ryu2002winding}. As the vector ($d_{x}$, $d_{z}$) is found to wind the origin twice when $|\lambda|<1$, the geometric interpretation suggests $\mathcal{W}_\text{n}=2$; in contrast, when $|\lambda|>1$, no complete cycle is observed, suggesting $\mathcal{W}_\text{n}=0$, as depicted on the right side of \rfig{Fig3}(c). We have also calculated $\mathcal{W}_\text{n}$ by following \req{Wn} and using the experimentally obtained values of $d_{x}$ and  $d_{z}$, with the results presented in \rfig{Fig3}(d) (the blue circles). The experimental results, while displaying certain fluctuations \red{and smeared transitions} due to the finite discretization of the BZ \red{and some random errors presented in measurements},  are in good agreement with the theoretical expectation (the pink line). \red{In the Supplementary Information, we show that by using a finer \rred{and nonuniform} discretization of the k-space, much sharper transitions can be observed}. In \rfig{Fig3}(d), we also present the evolution of $(P_{x},P_{y})$ and the contours for $P_{z}=0$. As $\lambda_{1}=\lambda_{2}=\lambda$ is set, the four removable vortices and antivortices of $(P_{x},P_{y})$ move in a symmetrically all-inward or all-outward way with the variation of $\lambda$, and the change of $\mathcal{W}_\text{n}$, or say topological phase transition,  matches well with the annihilation of them  at  $(0,0)$ for $\lambda=-1$ and $(\pi,\pi)$ for $\lambda=1$.

\red{It is worth noting that while the region with $\mathcal{W}_\text{n}=2$ covers the whole second-order topological
phase when $\lambda_{1}=\lambda_{2}=\lambda$, the number of zero-energy modes per corner in the resulting
second-order topological phase
is just $1$ rather than $2$~\cite{Yan2019HOTOPSC}. This
may look counterintuitive,  since it is known that the winding number is equivalent
to the number of zero-energy modes per end in one dimension. However, here
$\mathcal{W}_\text{n}$ is defined on a high symmetry line in two dimensions. Owing to the presence of an extra dimension,
what can be inferred from  $\mathcal{W}_\text{n}=2$ is the existence of gapless modes when the sample's edges are chosen along
the $x=y$ or $x=-y$ direction~\cite{Yan2019HOTOPSC}. On the other hand, the number of zero-energy mode at one corner is determined by two intersecting edges,
so there is no reason to expect that the simple relation between winding number and zero-energy mode
in one dimension should also
hold in two dimensions.}

\vspace{10pt}
\begin{spacing}{1.5}
\noindent\it{2.4 Topological phase transitions between second-order and first-order topological phases}
\end{spacing}
When  $\mu$ becomes variable, the number of BISs is no longer limited to be even, as a result, first-order topological phases become possible~\cite{sato2010odd} (see the phase diagram in \rfig{Fig4}(a)).  The first-order topological phase in \rfig{Fig4}(a) has Chern number $C=1$ as it has only one BIS within which the numbers of vortices and antivortices differ by one.

From \rfig{Fig4}(a), it is apparent that the desired topological phase transitions between second-order and first-order topological phases can be achieved by appropriately varying $\lambda$ or $\mu$. Without loss of generality,  we fix $\lambda=-0.5$ and vary $\mu$ in a broad regime (see the blue dashed line in \rfig{Fig4}(a)). Accordingly, the configuration of vortices and antivortices is expected to be intact and only the \red{BIS} will change. As here the first-order topological phase with $C=1$ has a corresponding $\mathcal{W}_\text{n}=1$, in the experiment we keep using the observed pseudo-spin texture  on the $k_{x}=k_{y}$ line to extract the topological invariants, as well as the topological phase transitions.

\red{As  a wide range of $\mu$ is to cover, here we choose a sparser discretization of the BZ. Concretely, we consider a uniform discretization with $31$ discrete momentum points in $k\in[-\pi,\pi]$.} Figure \ref{Fig4}(b) shows the pseudo-spin texture of $(d_{x},d_{z})$ extracted from the observed $(P_{x},P_{z})$. Also according to the number of cycles that the vector ($d_{x}$, $d_{z}$) winds around the origin, we find $\mathcal{W}_\text{n}=2$ for $-2<\mu<0$ (second-order \red{topological phase}), $\mathcal{W}_\text{n}=1$ for $0<\mu<4$ (first-order \red{topological phase}), and $\mathcal{W}_\text{n}=0$ otherwise (trivial \red{phase}). In \rfig{Fig4}(c), the $\mathcal{W}_\text{n}$ deduced experimentally using Eq. (\ref{Wn}) (the blue circles) is presented. \red{Because of a sparser discretization, it is readily seen that compared to
the topological phase transitions shown  in \rfig{Fig3}(d), the sharpness of the phase boundaries between
topologically distinct phases is relatively reduced}. Nevertheless, the experimentally obtained topological invariants  are still in good agreement with the theoretical expectation (the pink line). For the sake of completeness, we also show the pseudo-spin texture near the critical point ($\mu=0$) separating the second-order and first-order \red{topological phases} in \rfig{Fig4}(c).  One can see from the insets of \rfig{Fig4}(c) that when $\mu=-0.1$, $P_{z}=0$ has two disconnected contours; at the critical point $\mu=0$, the smaller one contour shrinks to a point and coincides with the time-reversal invariant momentum $(0,0)$, which leads to the closure of bulk gap; when $\mu=0.1$, the smaller one contour completely disappears and only the larger one remains. Both the evolution of pseudo-spin texture and the change of topological invariant confirm the realization of topological phase transitions between second-order and first-order \red{topological phases}.

%Before proceeding, we emphasize that the topological phase transitions between second-order TSCs and trivial superconductors and the ones between second-order and first-order TSCs  display fundamental difference. While the former can be achieved by simply annihilating the removable vortices and antivortices without the closure of bulk gap (see \rfig{Fig2}), the latter is inevitably associated with the closure of bulk gap as they involve the change of first-order topology

\begin{spacing}{2.0}
\noindent\textbf{3\quad Discussion and conclusion}
\end{spacing}

\red{Our experiment has demonstrated the feasibility of  the bulk approach to determine
the band topology. Although our experiment was performed in a single superconducting qubit
of great controllability, the basic idea is general and can be applied to other platforms
as long as there exist appropriate experimental methods to detect the underlying spin or pseudo-spin
texture of the bulk bands~\cite{Ji2020simulation,Xin2020simulation}.
For real materials, as aforementioned,  the complexity of the band structures and the fact that
the bulk topological invariants are determined by all occupied bands together
raise great challenge for the bulk approach to determine the underlying bulk topological invariants.
Nevertheless, it is justified to expect that even for real materials, the bulk approach
is still feasible for the detection of topological phase transitions
once powerful spin- and angle-resolved photoemission spectroscopy
(similar tools for other metamaterials) is developed. This is because
the  topological phase transitions in general only involve a few bands near the Fermi energy.
When the topological phase transitions belong to the $Z_{2}$ type, like
the transition between a three-dimensional inversion symmetric
strong topological insulator  and a trivial insulator,
the situation is further simplified since one only needs to
focus on a few high symmetry momenta and detect the evolution of
topological spin-texture configurations at their neighborhood.  }

\red{Another important message from our experiment is that flattening the Hamiltonian and a non-uniform sampling of the $k$-space can be very efficient techniques to extract the underlying pseudo-spin texture through the quantum-quench dynamics. As a precise determination of bulk topological invariants and phase boundaries generally requires dense measurements of the pseudo-spin texture throughout the \rred{BZ}, these techniques can considerably save the time and experimental resources, while still allowing us to explore a reasonably large parameter space. Such a merit will become more prominent when studying higher-dimensional topological phases, simply because the measurements required to determine the band topology will drastically increase as the dimension of \rred{BZ} increases. Therefore, we expect that these techniques would be widely adopted in the successive experimental studies of higher-dimensional topological phases and related topological phase transitions. }

In summary, we have simulated a second-order \red{topological phase} in a controllable superconducting qubit and observed novel topological phase transitions between this phase and a trivial \red{phase}, as well as a first-order \red{topological phase} within the same symmetry class.
Our work opens new opportunities for the experimental study of higher-order topological phases.  A direction forward is to investigate higher-order topological phases in higher dimensions, \red{where the higher-order topology can be determined by detecting the spin-texture winding of
 the so-called nested BISs
through the same approach as in our experiment~\rred{\cite{Yu2020BIS,Li2020HOTI}}}.
In addition, the nonequilibrium properties of quenched or driven higher-order topological phases are also of great interest and worthy of
in-depth studies.

\begin{acknowledgements}
This work was supported by the Key-Area Research and Development Program of Guang-Dong Province (Grant No. 2018B030326001), the National Natural Science Foundation of China (U1801661), the National Science Foundation of China (No.11904417), the Guangdong Innovative and Entrepreneurial Research Team Program (2016ZT06D348), the Guangdong Provincial Key Laboratory (Grant No.2019B121203002), the Natural Science Foundation of Guangdong Province (2017B030308003), and the Science, Technology and Innovation Commission of Shenzhen Municipality (JCYJ20170412152620376, KYTDPT20181011104202253), and the NSF of Beijing (Grants No. Z190012).
\end{acknowledgements}

%%%%%%%%%%%%%%%%%%%%%%%%%%%%%%%%%%%%%%%%%%%%%

\clearpage

%%%%%%%%%% Merge with supplemental materials %%%%%%%%%%
\pagebreak
\widetext
%\begin{center}
%\textbf{\large Supplemental Material}
%\end{center}
\section*{\normalsize SUPPLEMENTAL MATERIAL}

%%%%%%%%%% Merge with supplemental materials %%%%%%%%%%

%%%%%%%%%% Prefix a "S" to all equations, figures, tables and reset the counter %%%%%%%%%%
\setcounter{equation}{0}
\setcounter{figure}{0}
\setcounter{table}{0}
\setcounter{page}{1}
\makeatletter
\renewcommand{\theequation}{S\arabic{equation}}
\renewcommand{\thefigure}{S\arabic{figure}}
\renewcommand{\bibnumfmt}[1]{[S#1]}
\renewcommand{\citenumfont}[1]{S#1}
%%%%%%%%%% Prefix a "S" to all equations, figures, tables and reset the counter %%%%%%%%%%

This supplemental information contains the following sections: (I) The derivation of the formula for time-averaged pseudo-spin polarization; (II) The determination of phase diagram; (III) Pseudo-spin polarization determined in the experiment;
(IV) Removable Dirac points and  topological phase
transitions between second-order \red{topological phases}  and trivial \red{phases}; (V) The change of pseudo-spin texture across topological phase transitions; \red{ (VI) The scheme of Brillouin zone discretization and the sharpness of the phase boundaries between
topologically distinct phases;} (VII) Information of samples and experimental setup.
\vspace{-8pt}

\section{I. Time-averaged pseudo-spin polarization}

In the experiment, we start with the trivial ground state $|\psi(\mathbf{k},t)\rangle$
corresponding to the limiting situation $\lambda_{1,2}\rightarrow\infty$. That is,
$|\psi(\mathbf{k},t)\rangle$ is an eigenstate of $\sigma_{z}$, i.e.,
$\sigma_{z}|\psi(\mathbf{k},t)\rangle=-|\psi(\mathbf{k},t)\rangle$.
At a time (we take it as the reference time $t=0$), we suddenly quench the system by a microwave
pulse, then the state will follow a unitary time evolution, with
\begin{eqnarray}
|\psi_{f}(\mathbf{k},t)\rangle=\mathcal{T}e^{-i\int_{0}^{t}H_{f}(\mathbf{k})dt}|\psi(\mathbf{k},0)\rangle,
\end{eqnarray}
where $\mathcal{T}$ stands for time-ordering operator and
$H_{f}(\mathbf{k})=\mathbf{d}(\mathbf{k})\cdot\boldsymbol{\sigma}$ takes the form to simulate.

After the quench, the pseudo-spin polarization, which is
defined as $\mathbf{P}(\mathbf{k},t)=\langle
\psi_{f}(\mathbf{k},t)|\boldsymbol{\sigma}|\psi_{f}(\mathbf{k},t)\rangle$,
evolves with time. Below we show when the spectra of $H_{f}(\mathbf{k})$ are flatten, that is
$H_{f}(\mathbf{k})\rightarrow \bar{H}(\mathbf{k})=H_{f}(\mathbf{k})/E_{f}(\mathbf{k})$, where
$E_{f}(\mathbf{k})=d(\mathbf{k})=\sqrt{d_{x}^{2}(\mathbf{k})+d_{y}^{2}(\mathbf{k})+d_{z}^{2}(\mathbf{k})}$,
the pseudo-spin polarization becomes time periodic and its average value over one period has a simple
relation with the $\mathbf{d}$ vector.

For the flattened Hamiltonian, we have
\begin{eqnarray}
|\psi_{f}(\mathbf{k},t)\rangle&=&\mathcal{T}e^{-i\int_{0}^{t}\bar{H}(\mathbf{k})dt}|\psi(\mathbf{k},0)\rangle\nonumber\\
&=&(\cos t-i\sin t
\frac{\mathbf{d}(\mathbf{k})\cdot\boldsymbol{\sigma}}{d(\mathbf{k})})|\psi(\mathbf{k},0)\rangle,
\end{eqnarray}
then
\begin{eqnarray}
\mathbf{P}(\mathbf{k},t)&=&\langle\psi(\mathbf{k},0)|(\cos t+i\sin t
\frac{\mathbf{d}(\mathbf{k})\cdot\boldsymbol{\sigma}}{d(\mathbf{k})})
\boldsymbol{\sigma}(\cos t-i\sin t
\frac{\mathbf{d}(\mathbf{k})\cdot\boldsymbol{\sigma}}{d(\mathbf{k})})|\psi(\mathbf{k},0)\rangle\nonumber\\
&&=\cos^{2} t\langle\psi(\mathbf{k},0)|\boldsymbol{\sigma}
|\psi(\mathbf{k},0)\rangle+\sin^{2}t\langle\psi(\mathbf{k},0)|
\frac{\mathbf{d}(\mathbf{k})\cdot\boldsymbol{\sigma}}{d(\mathbf{k})}
\boldsymbol{\sigma}\frac{\mathbf{d}(\mathbf{k})\cdot\boldsymbol{\sigma}}{d(\mathbf{k})}|\psi(\mathbf{k},0)\rangle\nonumber\\
&&
+i\cos t\sin t\langle\psi(\mathbf{k},0)|
[\frac{\mathbf{d}(\mathbf{k})\cdot\boldsymbol{\sigma}}{d(\mathbf{k})},\boldsymbol{\sigma}]|\psi(\mathbf{k},0)\rangle.
\end{eqnarray}
The expression for time-averaged pseudo-spin polarization in Eq.(4) of the
main text is obtained by
\begin{eqnarray}
\mathbf{P}(\mathbf{k})&=&\frac{1}{T}\int_{0}^{T}\mathbf{P}(\mathbf{k},t)dt\nonumber\\
&=&\frac{1}{2}\langle\psi(\mathbf{k},0)|\boldsymbol{\sigma}
|\psi(\mathbf{k},0)\rangle+\frac{1}{2}\langle\psi(\mathbf{k},0)|
\frac{\mathbf{d}(\mathbf{k})\cdot\boldsymbol{\sigma}}{d(\mathbf{k})}
\boldsymbol{\sigma}\frac{\mathbf{d}(\mathbf{k})\cdot\boldsymbol{\sigma}}{d(\mathbf{k})}|\psi(\mathbf{k},0)\rangle.
\end{eqnarray}
By using $\langle\psi(\mathbf{k},0)|\sigma_{z}|\psi(\mathbf{k},0)\rangle=-1$ and
$\langle\psi(\mathbf{k},0)|\sigma_{x,y}|\psi(\mathbf{k},0)\rangle=0$, a further step leads to the
final expression
\begin{eqnarray}
P_{i}(\mathbf{k})=-\frac{d_{i}(\mathbf{k})d_{z}(\mathbf{k})}{d_{x}^{2}(\mathbf{k})+d_{y}^{2}(\mathbf{k})+d_{z}^{2}(\mathbf{k})}.
\end{eqnarray}

\section{II. The determination of phase diagram}

Here we provide the details about the determination of the $\lambda$-$\mu$ phase diagram.
Considering $\lambda_{1}=\lambda_{2}=\lambda$, the three components of the $\mathbf{d}$
vector are given by
\begin{eqnarray}
d_{x}(\mathbf{k})&=&2\lambda(\sin k_x +  \sin k_y) + \sin 2k_x +\sin 2k_y,\nonumber\\
d_{y}(\mathbf{k})&=&2\lambda(\sin k_y -  \sin k_x) + 2\sin (k_y-k_x),\nonumber\\
d_{z}(\mathbf{k})&=&2\lambda(\cos k_x + \cos k_y) + \cos 2k_x +\cos 2k_y-\mu.
\end{eqnarray}
\red{For the convenience of discussion, we name the contours satisfying $d_{z}(\mathbf{k})=0$
as band inversion surfaces (BISs) and the points simultaneously satisfying $d_{x}(\mathbf{k})=d_{y}(\mathbf{k})=0$
as Dirac points (DPs)}. For this Hamiltonian, the change of first-order topology is associated with
the change of the parity of \red{the number of BISs ($N_{\rm BIS}$)}. One can readily find that
it takes place at the two time-reversal invariant momenta, $(0,0)$ and $(\pi,\pi)$.
Therefore, there are two critical $\mu$ for the topological phase
transitions between first-order topological phases and other phases. The critical $\mu$ should lead
$d_{z}$ to satisfy
$d_{z}(0,0)=0$ or $d_{z}(\pi,\pi)=0$. Accordingly, we find
\begin{eqnarray}
\mu_{c,1}=4\lambda+2, \qquad
\mu_{c,2}=-4\lambda+2.
\end{eqnarray}
As there is only one \red{BIS} when
$\mu\in(-4|\lambda|+2, 4|\lambda|+2)$, the regime $(-4|\lambda|+2, 4|\lambda|+2)$
corresponds to the first-order topological phase.

\begin{figure}[htbp]
\includegraphics[width=0.45\textwidth]{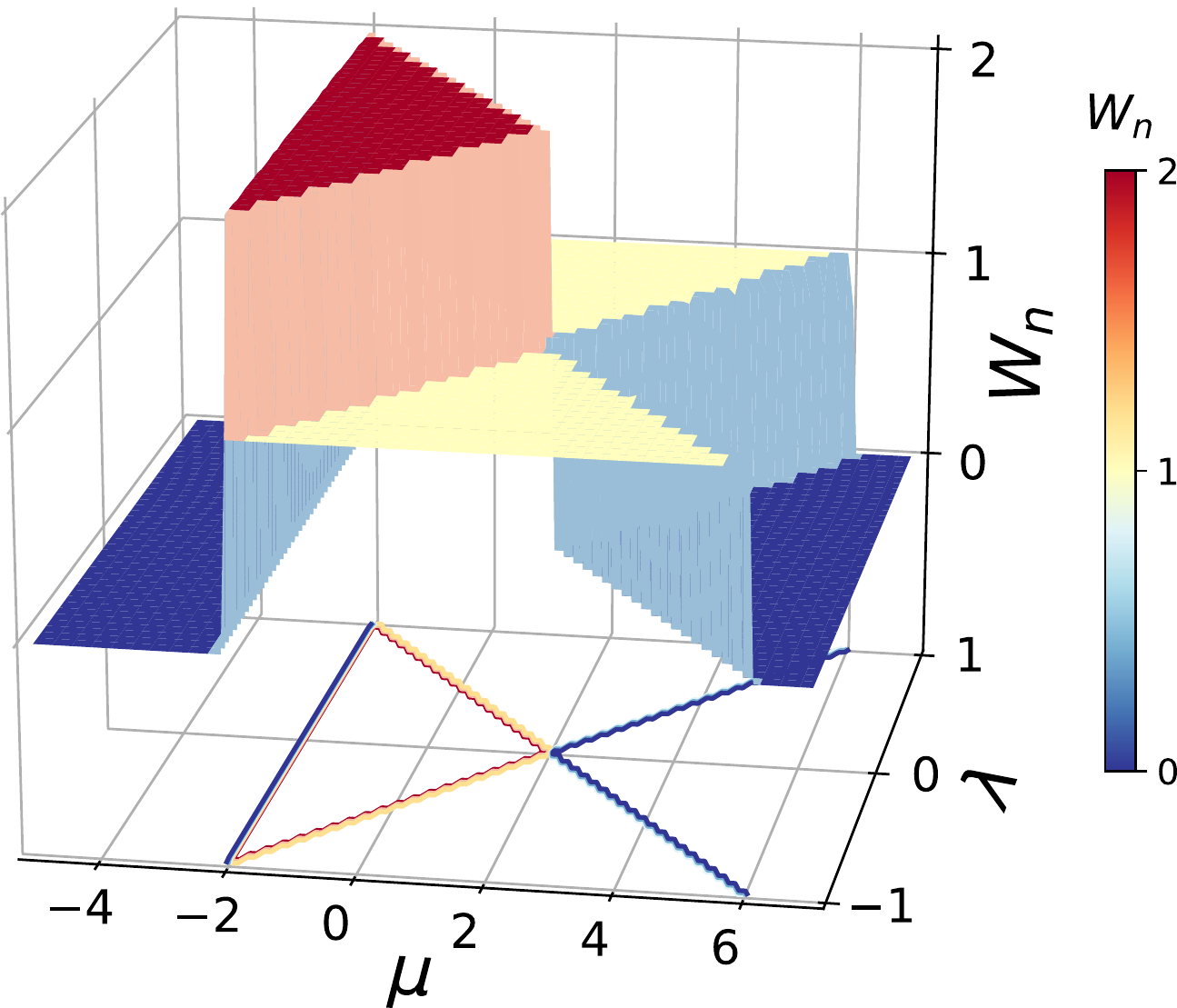}
\centering
\caption{The calculated winding number $\mathcal{W}_\text{n}$ with varying $\lambda$ and $\mu$.}
\label{figs1}
\end{figure}

When the parity of \red{$N_{\rm BIS}$ is even and one of the BISs}
crosses the four removable \red{DPs} (symmetry enforces that the crossing takes
place simultaneously for the four removable \red{DPs}), the system undergoes
a topological phase transition between second-order topological phases and other
phases.
As $d_{x}(\mathbf{k})-id_{y}(\mathbf{k})=2[(\cos k_{x}+\lambda)+i(\cos
k_{y}+\lambda)]
(\sin k_{x}-i\sin k_{y})$, it is readily found that the four removable \red{DPs}
are located at $(k_{x;\pm},k_{y,\pm})=(\pm(\pi-\arccos \lambda),\pm(\pi-\arccos \lambda))$.
At these four momenta, $d_{z}(k_{x;\pm},k_{y,\pm})=-2-\mu$, and we obtain another critical
 value for $\mu$ when $|\lambda|<1$, which is
\begin{eqnarray}
\mu_{c,3}=-2.
\end{eqnarray}
The three phase boundaries divide the phase diagram into three topologically distinct regimes,
as shown in \rfig{figs1}.

Let us now focus on the two high symmetry lines with $k_{x}=\pm k_{y}$, on which
the Hamiltonian is reduced to
\begin{eqnarray}
H(k_{+})&=&d_{x}(k_{+})\sigma_{x}+d_{z}(k_{+})\sigma_{z}, \nonumber\\
H(k_{-})&=&d_{y}(k_{-})\sigma_{y}+d_{z}(k_{-})\sigma_{z},
\end{eqnarray}
where $k_{+(-)}$ represents the momentum on the line $k_{x}=(-) k_{y}$. As $\{H(k_{+}),\sigma_{y}\}=0$
and $\{H(k_{-}),\sigma_{x}\}=0$,
the two reduced Hamiltonians both have chiral symmetry so their topological properties
are characterized by the winding number, which can be written down compactly as
\begin{eqnarray}
\mathcal{W}_{n;\pm}=\frac{1}{4\pi
i}\int_{-\pi}^{\pi}\text{Tr}[C_{\pm}H^{-1}(k_{\pm})\frac{\partial}{\partial
k_{\pm}}H(k_{\pm})]dk_{\pm},
\end{eqnarray}
where $C_{+}=\sigma_{y}$ and $C_{-}=\sigma_{x}$ represent the respective chiral operators. If
in terms of the $\mathbf{d}$ vector, we have
\begin{eqnarray}
\mathcal{W}_{n;+}&=&\frac{1}{2\pi}\int_{-\pi}^{\pi} \frac{d_z\partial_{k_{+}} d_x - d_x
\partial_{k_{+}} d_z}{d_x^2+d_z^2}\, dk_{+},\label{winding}\\
\mathcal{W}_{n;-}&=&\frac{1}{2\pi}\int_{-\pi}^{\pi} \frac{d_y\partial_{k_{-}} d_z - d_z
\partial_{k_{-}} d_y}{d_y^2+d_z^2}\, dk_{-}.
\end{eqnarray}
Depending on the choice $k_{-}=k_{x}=-k_{y}$ or $k_{-}=k_{y}=-k_{x}$, we have
either  $\mathcal{W}_{n;+}=\mathcal{W}_{n;-}$ or $\mathcal{W}_{n;+}=-\mathcal{W}_{n;-}$.
As only the absolute value of  winding number does not depend on the choice,
throughout the whole paper, we only care about the absolute value.
A simple numerical calculation reveals that $\mathcal{W}_{n;\pm}=1$
for the first-order topological phase,
$\mathcal{W}_{n;\pm}=2$ for the second-order topological  phase, and
$\mathcal{W}_{n;\pm}=0$ for the trivial phase, suggesting that
the winding number can fully distinguish all phases from each other.

\begin{figure*}[htbp]
	\begin{center}
		\includegraphics[width=0.6\textwidth]{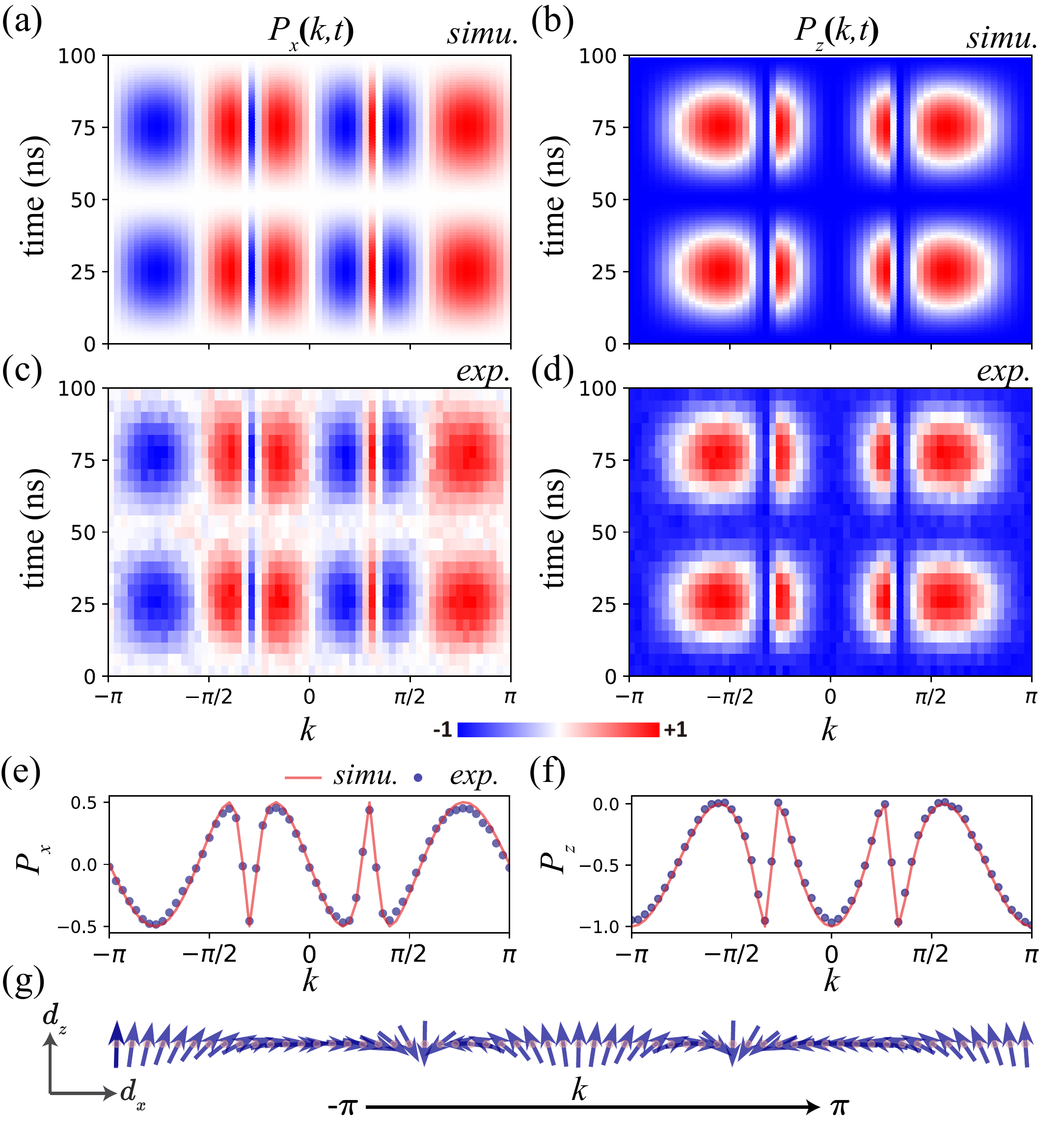}
		\caption{Time-evolution of the pseudo-spin polarization $P_{x,z}(\mathbf{k},t)$, for a typical
example of $\lambda=-0.5$, $\mu=-1.5$. The numerical and experimental results are displayed in (a)(b)
and (c)(d), respectively. (e-f) The corresponding time-averaged pseudo-spin polarization $P_{x,z}$. The solid lines show the simulation results, and the symbols are the experimental data. The error in the experimental data is in a range between 1-2\% (see the text following for an explanation of how the error is estimated), which is too small to be depicted by error bars. (g) The two-component vector $(d_x, d_z)$ along $k_x=k_y=k$ line experimentally deduced from (e-f). }
		\label{figs2}
	\end{center}
\end{figure*}

\section{III. Pseudo-spin polarization determined in the experiment}

To demonstrate that quantum state tomography (QST) can faithfully determine the pseudo-spin polarization, in
this part we show the pseudo-spin polarizations
measured in the experiment and those predicted according to the theoretical model
together for comparison. For the sake of concreteness, we consider
a sudden quench from $\lambda \rightarrow \infty$ (trivial) to $\lambda=-0.5$ and $\mu=-1.5$
(second-order \red{topological phase}) and focus on the pseudo-spin polarization on the high symmetry line
$k_{x}=k_{y}$, on which $d_{y}=0$ and so the $y$-component of the
time-averaged pseudo-spin polarization is zero. As we are only interested in the time-averaged pseudo-spin polarization, in this part we will only show the evolution of $P_{x}(\mathbf{k},t)$ and $P_{z}(\mathbf{k},t)$ whose time average are
expected to be nonzero.
Fig.\ref{figs2}(a)(b) show the evolution of the
$x$- and $z$-components of the pseudo-spin polarization with time predicted by theory, and Fig.\ref{figs2}(c)(d) show the respective evolutions measured by QST. It is readily seen that the experimental results agree very well with the predicted values. Fig.\ref{figs2}(c)(d) also confirm the periodic behavior of pseudo-spin polarization.

According to the measured experimental data, we present the time-averaged pseudo-spin polarization in Fig.\ref{figs2}(e)(f). The solid symbols are experimental results deduced from Fig.\ref{figs2}(c)(d). We estimate the error in the experimental data in the following way. Prior to the measurement of each time-evolution curve (for example, in Fig.\ref{figs2}(c), one such curve corresponds to a fixed value of $k$ in a time range between 0 and 100 ns), we perform a set of 30 repeated measurements to calibrate the preparation of the states of $|0\rangle$ and $|1\rangle$. The standard error of the results of such measurements is taken as the error of the following measurement of the corresponding time-evolution curve. $d_{x}$ and $d_{z}$ can then be extracted from the experimental results in Fig.\ref{figs2}(e)(f). Fig.\ref{figs2}(g) shows the winding behavior of the two-component vector ($d_{x}$, $d_{z}$) across the Brillouin zone. It is easy to see that the two-component vector undergoes two complete cycles of winding when $k$ goes from $-\pi$ and $\pi$, suggesting $\mathcal{W}_{n}=2$ and so the realization of the second-order topological phase.

\begin{figure*}[htbp]
	\begin{center}
		\includegraphics[width=0.95\textwidth]{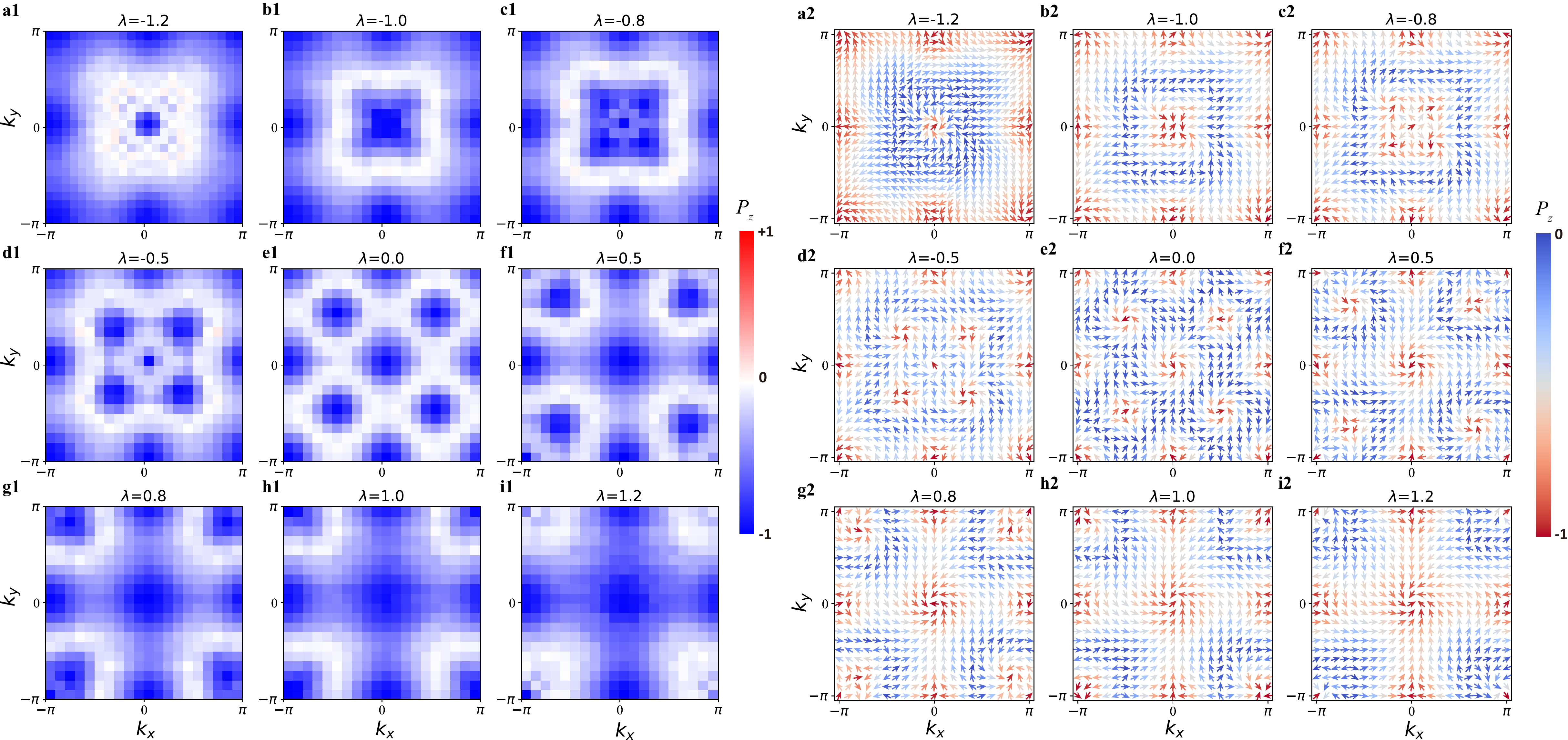}
		\caption{$\lambda$-dependent removable \red{Dirac points} and \red{band inversion surfaces}. (a)-(i) The \red{band inversion surfaces} (left) and pseudo-spin textures (right) displayed by $P_z$ and ($P_x$, $P_y$), respectively, for a wide range of
$\lambda$ from $\lambda<-1$ to $\lambda>1$.}
		\label{figs3}
	\end{center}
\end{figure*}

\section{IV. Removable Dirac points and
topological phase transitions between second-order topological phases and trivial
phases}

In this part, we provide more experimental data for the case with $\mu=-\lambda_{1}^{2}-\lambda_{2}^{2}$. As mentioned in the main text, because of the constrain from the Hopf map, this case has only two topologically distinct phases, the second-order topological phase for $|\lambda_{1,2}|<1$, and the trivial  phase otherwise.

Focusing on $\lambda_{1}=\lambda_{2}=\lambda$, we tune $\lambda$ from $-1.2$ to $1.2$. In \rfig{figs3}, (a1)-(i1) on the left panel show the evolution of $P_{z}$, and (a2)-(i2) on the right panel show the evolution of ($P_x$, $P_y$). It is readily seen that when $\lambda$ goes across $\lambda_{c,l}=-1$ from below, two additional vortices and two additional antivortices emerge. With the further increase of $\lambda$,  their positions move away from the time-reversal invariant momentum $(0,0)$ in a symmetric way, while remaining located between the two contours for $P_{z}=0$. When $\lambda$ reaches $\lambda_{c,u}=1$, the four movable vortices and antivortices coincide at the time-reversal invariant momentum $(\pi,\pi)$ and then annihilate with each other. At $\lambda=\pm 1$, it seems that there is only one contour satisfying $P_{z}=0$ (see (b1) and (h1)). In fact, this corresponds to the critical situation for which one of the contour is shrunk to a zero-size point (the time-reversal invariant momentum $(0,0)$ for $\lambda=-1$ and $(\pi,\pi)$ for $\lambda=1$).

\begin{figure}[htbp]
\includegraphics[width=0.6\textwidth]{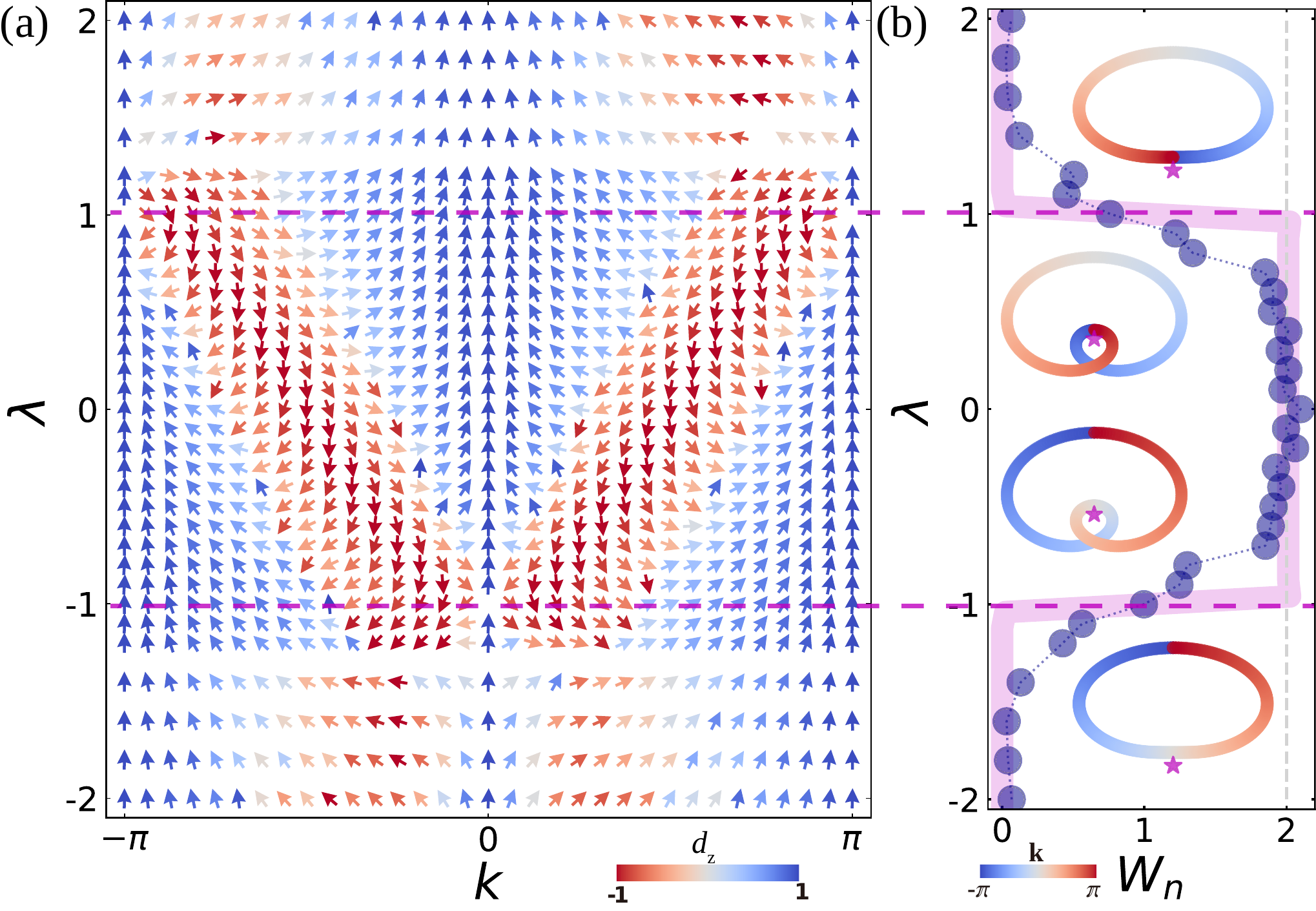}
\centering
\caption{Pseudo-spin texture ($d_y$, $d_z$) (a) and deduced winding number $\mathcal{W}_{n}$ (b) along
the $k_x=-k_y$ line.}
\label{figs4}
\end{figure}

In \rfig{figs4}, we show the pseudo-spin texture on the high symmetry line $k_{x}=-k_{y}$.
As $d_{x}$ vanishes on this line, the winding number is determined by the structure of
$(d_{y},d_{z})$. As expected, the winding number is found to be $\mathcal{W}_{n}=2$ in the regime
$-1<\lambda<1$, and $\mathcal{W}_{n}=0$ when $\lambda<-1$ or $\lambda>1$, suggesting that the presence
of removable vortices (or say \red{DPs}) is crucial for the realization
of second-order topological phase.

\begin{figure}[htbp]
\includegraphics[width=0.6\textwidth]{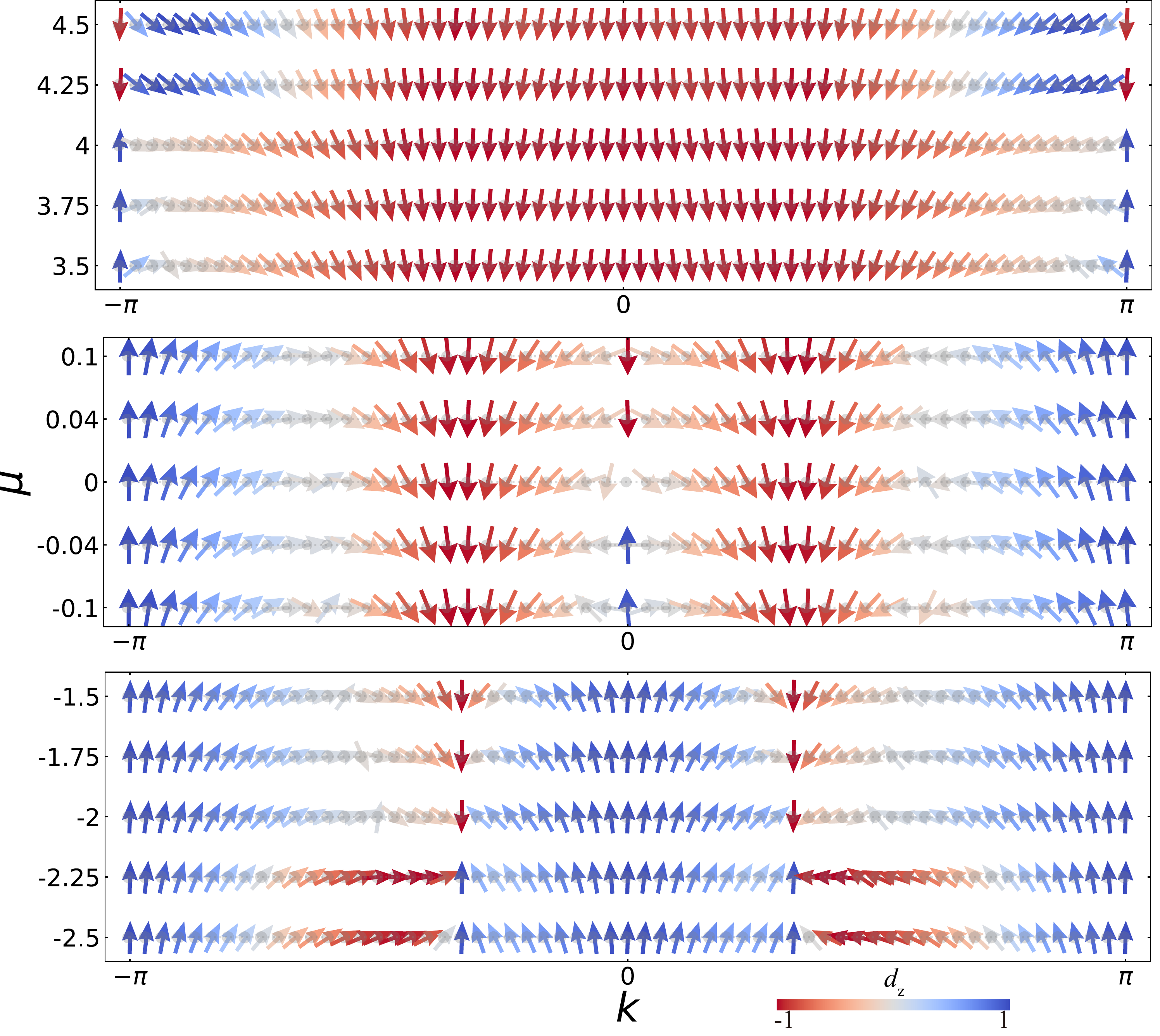}
\centering
\caption{A detailed display of ($d_x$, $d_z$) near the phase boundaries around $\mu=-2,0,4$
respectively.}
\label{figs5}
\end{figure}

\section{V. The change of pseudo-spin texture across topological phase transitions}

In this part, we provide a better view of the change of pseudo-spin texture across
topological phase transitions. Here we still consider the case discussed in the main text.
That is $\lambda=-0.5$. For this case, the critical points correspond to
$\mu=-2$, $0$ (that is $4\lambda+2$), $4$ (that is $-4\lambda+2$). The pseudo-spin textures near these
three critical points are presented
in \rfig{figs5}. From the results, it is readily found that
when $\mu<-2$, the two-component vector $(d_{x},d_{z})$ does not wind
a complete cycle when $k$ ($k=k_{x}=k_{y}$) goes from $-\pi$ to $\pi$, suggesting
$\mathcal{W}_{n}=0$; when $-2<\mu<0$,
the two-component vector $(d_{x},d_{z})$ is found to be able to wind the cycle completely
when $k$ goes from $-\pi$ to $\pi$, and
the time of complete-cycle winding is found to be $2$, suggesting $\mathcal{W}_{n}=2$;
when $0<\mu<4$,  the two-component vector $(d_{x},d_{z})$ is also found to be able to wind the cycle
completely, but the time of complete-cycle winding is changed to $1$, suggesting $\mathcal{W}_{n}=1$;
when $\mu>4$, again the two-component vector $(d_{x},d_{z})$ does not wind
a complete cycle, suggesting that the system returns the trivial  phase with
$\mathcal{W}_{n}=0$.

\red{\section{VI. The scheme of Brillouin zone discretization and the sharpness of the phase boundaries between
topologically distinct phases}}
\red{ Above we have shown that the time of complete-cycle winding provides an intuitive geometric picture to extract the winding number.
To determine the winding number more strictly, we need to follow its algebraic expression. Let us still focus on the high symmetry line $k_{x}=k_{y}=k$ for illustration, accordingly, the winding number is given by Eq.(\ref{winding}) or Eq.(4) in the main text. Theoretically, to obtain the quantized value of the winding number, one needs to
perform a continuous measurement of the pseudo-spin texture from $k=-\pi$ to $\pi$. However, this is apparently unrealistic in any real experiment
due to the consumption in time and experimental resources.
\begin{figure}[htbp]
\includegraphics[width=10cm, height=8cm]{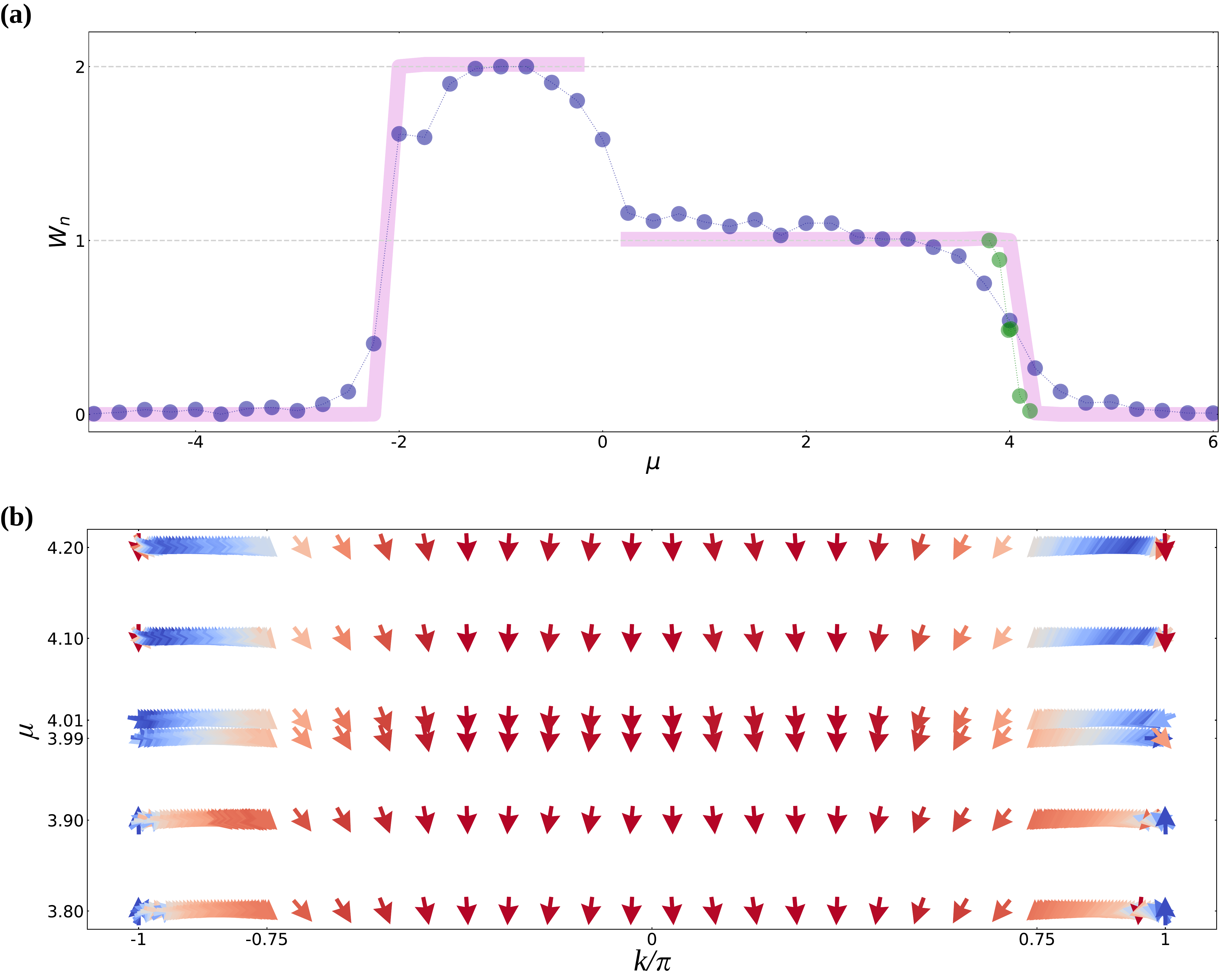}
\centering
\caption{\red{ Experimentally measured winding numbers and pseudo-spin textures.
(a) The purple dots correspond to winding numbers which are obtained by
a uniform discretization of the Brillouin zone to $31$ points (these data are
taken from Fig.4(c) of the main text for comparison). As a comparison, we have measured six new groups of pseudo-spin textures near
the critical point at $\mu=4$. These six groups of pseudo-spin textures, which lead to the six winding numbers labeled by greed dots,
  are measured under
a nonuniform discretization of the Brillouin zone to $102$ points.
(b) shows our scheme of the momentum discretization for the six green dots in (a) and the corresponding
pseudo-spin textures at the chosen discrete momenta.
Within the region considered for illustration, i.e., $\mu\in[3.8,4.2]$, the pseudo-spin texture near $k=\pi$ changes strongly
with the variation of $\mu$,
while the pseudo-spin texture near $k=0$ does not change much. Therefore,  we perform intensive measurements near $k=\pi$ ($41$ momentum points are considered both
in $[-\pi, -0.75\pi]$ and in  $[0.75\pi, \pi]$) while
measuring the pseudo-spin texture near $k=0$ only coarsely (only $20$ momentum points are considered within $[-0.75\pi, 0.75\pi]$). The winding numbers for
the six green dots in (a) clearly show that a finer discretization
of the momentum can determine the phase boundary much more sharply.}
}  \label{figs6}
\end{figure}
In the experiment, we have to discretize the Brillouin zone. Due to the same reason, the changes of $\lambda$ and $\mu$ have also to be discretized.
Roughly estimating, to obtain the results in Fig.3 and Fig.4 of the main text,
the experimental time for measuring pseudo-spin textures is $T=N_{p}\times N_{m}\times t$,
where $N_{p}$ denotes the number of discrete values for  $\lambda$ or $\mu$
under investigation, $N_{m}$ denotes the number of discrete momentum points at which
the pseudo-spin polarizations are measured, and $t$ denotes the time required for each discrete momentum point. In our experiment,
the technique of flattening the Hamiltonian has greatly reduced the measurement time required
for achieving a given precision, compared to the case of mapping the original Hamiltonian.  As $t$ is almost
fixed, what needs to be balanced is  $N_{p}$ and $N_{m}$. Although  an increase of $N_{m}$ will naturally reduce the error induced by the
discretization of Brilloin zone, it also considerably increases the total time (experimental resources as well)
if $N_{p}$ is fixed. }

\red{In the experiment, we have adopted a uniform discretization. Concretely,  we have chosen $\{N_{p},N_{m}\}=\{41,41\}$ and $\{45,31\}$
in Fig.3 and Fig.4, respectively. In the deep trivial regime, one can infer from Fig.3(d) and Fig.4(c) that under these discretizations,
the momentum points are already dense enough
to reach an excellent agreement between theory and experiment. The underlying reason  is that
 the pseudo-spin polarization changes rather smoothly with respect to momentum
 in the deep trivial regime, so even a relatively sparse discretization can lead
to an excellent agreement. Also from  Fig.3(d) and Fig.4(c), one can find that the phase boundaries separating topologically
distinct phases are not
very sharp. The underlying reason is that when getting close to these phase boundaries, the pseudo-spin polarization will change drastically
near the critical momenta at which the energy gap gets closed, consequently resulting in a considerable discrepancy between theory and experiment if the
discretizations in these regions are sparse.  To enhance the sharpness, a finer discretization at the neighborhood of these critical momenta is required. Therefore, when getting close to the critical points, it is better to adopt a nonuniform
discretization in the experiment. Near the critical momenta, measurements must be performed with a smaller step size in momentum. Accordingly, one should also allocate more measurement time (e.g., to increase the number of averages) to such regions where experimental errors (the major experimental error exists in the measurement of the pseudo-spin polarization) have a much more prominent effect. Away from the critical momenta, the step size in momentum can be chosen larger as the pseudo-spin polarization changes smoothly. Here we take the phase boundary at $\mu=4$ for example. As discussed in Sec.II, when $\mu=4$, $d_{z}$ takes zero value at $k=\pi$. As $d_{x}$
and $d_{y}$ identically vanish at this momentum,  the energy gap thus also gets closed at this momentum. Accordingly, we adopt a
nonuniform discretization,  with 41 momentum points in $k\in[-\pi, -0.75\pi]$, 20 momentum points in  $k\in[-0.75\pi, 0.75\pi]$, and
41 momentum points in $k\in[0.75\pi, \pi]$. Compared to the previous discretization with 31 momentum points
from $-\pi$ to $\pi$ in total, the sharpness of the phase boundary at
$\mu=4$ is greatly enhanced, as shown by the green dots in Fig.\ref{figs6}(a). Fig.\ref{figs6}(b) shows the
corresponding pseudo-spin textures which are the basis to obtain the six winding numbers labeled by the greed dots
in Fig.\ref{figs6}(a). }

%\vspace{10pt}

\section{VII. Information of samples and experimental setup}
Our experiments were carried out using cross-shaped Xmon type of superconducting qubits. The samples were fabricated based on Al/AlO$_x$/Al Josephson junctions on sapphire substrates. The superconducting transition temperature for Al is about 1.2 K, and the base temperature of the dilution refrigerator is about 20 mK, which is sufficiently low for our devices.
\begin{figure}[htbp]
\includegraphics[width=0.9\textwidth]{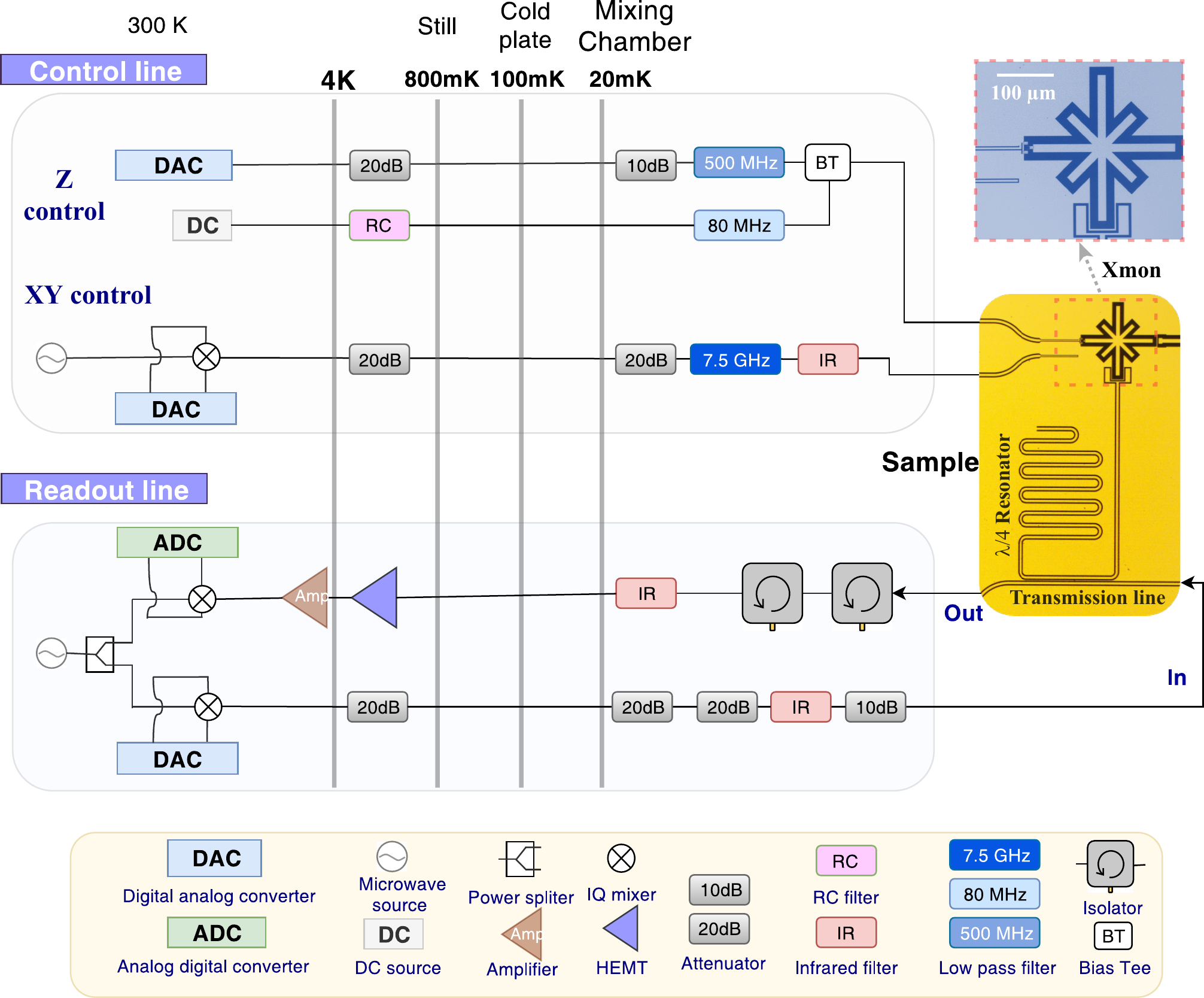}
\centering
\caption{Experimental setup. Including wiring, circuit components, as well as the sample image. }
\label{setup}
\end{figure}
Fig.\ref{setup} is the schematic of our experimental setup, showing both wiring scheme and relevant components from room temperature down to the base temperature stages. The samples are mounted in sample boxes made of Al that supply electromagnetic shielding at low temperatures.

The yellow- and blue-shaded areas on the right of Fig.\ref{setup} are optical images of one Xmon qubit used in our experiments. Three arms of the cross-shaped capacitance (blue-shaded image) are connected to different elements for coupling to other qubits (right, irrelevant for the current experiment), XY- and Z-control (left), and readout (bottom), respectively. The frequency of the qubit, $\omega_{01}$, is set by the bias on the Z-control line. Microwave pulses in the form of $\Omega(t)=\Omega \cos((\omega_{01}+\delta)t+\phi)$ are applied via the XY-control line for various qubit operations. The Xmon qubit is dispersively coupled to a $\lambda$/4 resonator (with a characteristic frequency of $\omega_{r} /2\pi= 6.8308~ \mathrm{GHz}$) for readout. More details regarding the design and control of the Xmon-type of qubits, as well as the dispersive readout technique, can be found in \cite{Barends2014S,Barends2013S,Kelly2015S,Reed2010S,Sank2016S}.

All data of this work, except for those presented in \rfig{figs4}, were obtained from one superconducting Xmon qubit. The qubit frequency is $f_{10}=\omega_{01}/2\pi = 5.04$ GHz, and the relaxation and dephasing times at this frequency are $T_1=17.48$ $\mu$s and  $T_2=0.82$ $\mu$s, respectively. The data in \rfig{figs4} were acquired on a different qubit with the following parameters: $f_{10}=\omega_{01}/2\pi = 5.908$ GHz, $T_1=3.81$ $\mu$s and  $T_2=1.84$ $\mu$s.

\section*{SUPPLEMENTARY REFERENCES}

%\bibliography{HOTSC}
%\bibliographystyle{myscibull}

%%%%%%%%%%%%%%%%%%%%%%%%%%%%%%%%%%%%%%%%%%%%%

\end{document}